# Topological quantum materials from the viewpoint of chemistry

Nitesh Kumar[‡], Satya N. Guin, Kaustuv Manna, Chandra Shekhar, and Claudia Felser[*]

Max Planck Institute for Chemical Physics of Solids, 01187 Dresden, Germany

**Abstract**: Topology, a mathematical concept, has recently become a popular and truly transdisciplinary topic encompassing condensed matter physics, solid state chemistry, and materials science. Since there is a direct connection between real space, namely atoms, valence electrons, bonds and orbitals, and reciprocal space, namely bands and Fermi surfaces, via symmetry and topology, classifying topological materials within a single-particle picture is possible. Currently, most materials are classified as trivial insulators, semimetals and metals, or as topological insulators, Dirac and Weyl nodal-line semimetals, and topological metals. The key ingredients for topology are: certain symmetries, the inert pair effect of the outer electrons leading to inversion of the conduction and valence bands, and spin-orbit coupling. This review presents the topological concepts related to solids from the viewpoint of a solid-state chemist, summarizes techniques for growing single crystals, and describes basic physical property measurement techniques to characterize topological materials beyond their structure and provide examples of such materials. Finally, a brief outlook on the impact of topology in other areas of chemistry is provided at the end of the article.

## Contents

## 1. Introduction

We use a variety of synthetic and natural solid materials in our daily lives. Recently, solids have been reclassified through the lens of topology, which goes far beyond the simple sum of their symmetry elements. All known inorganic compounds have been categorized using a single electron approach into trivial and topological materials[1-4] and published on the web.[5,6] All scientists can now search for new topological compounds on these web pages. This new viewpoint has led to the discovery of many unexpected properties, that include large responses to external stimuli, such as field (electric and magnetic), to waves (from light to acoustic waves), and to temperature, pressure, strain *etc* (Figure 1).[7-12] So far, we believe that this is only the tip of the iceberg. To date, physicists have mostly contributed to the successful story of topology. A number of solid-state chemists, particularly those who have been influenced by the works of Roald Hoffman[13,14], have joined the topological community.[15-18] Now is the time for topology to consider new avenues beyond those of condensed matter physics for example, for catalysis, and solar cells and beyond.

Another important and perhaps surprising outcome of research on topology in condensed matter physics is the prediction and realization of table-top experiments for high-energy physics and astrophysics.[9] So-called quasiparticles (electrons, holes, phonons *etc.*) in topological materials can mimic high energy particles such as the axion[19,20] or Majorana particle[21-24] or fields in the universe[9].

Topology can have impact in many areas of materials research and solid-state chemistry. In compounds such as Dirac semimetals (graphene being the first example), and Weyl semimetals, giant mobilities[8,25], small thermal and hydrodynamic electric conductivities[26], large chiral photocurrents[10], giant magnetoresistance[7,8] and Nernst effects[27,28] have been observed, with strong violation of "classical laws" such as the Wiedemann-Franz law[26], a law which limits the figure of merit of thermoelectric materials[29]. In magnetic Weyl semimetals large anomalous Hall[30-37], anomalous Nernst[38-40] and magneto optic effects[41] have been predicted or measured recently. Redox catalysis may also profit from topological properties[42] such as topological protected surface states[43-45], chiral surface states[46], giant electron mobilities[47,48], despite insulating and semiconducting bulk electronic structures. In the second part of our review, we discuss several examples of giant responses in greater detail to stimulate further research by chemistry groups.

The basic ingredient needed for most topological effects are relativistic effects and, therefore, it is not surprising that in many compounds heavy elements are important building blocks. Relativity contributes in two ways to topological materials, via the inert pair effect, and via spin-orbit coupling (SOC). The inert pair effect is responsible for lowering the energy of the outer *s* electrons 5*s* and 6*s* due to the nearly relativistic speed of the 1*s* electrons in heavy elements such as gold[49] and bismuth[16]. Many of the heavy *p*-block elements, such as bismuth, consequently have a different electron configuration in compounds compared to their lighter relatives. In ionic compounds such as $Bi_2Se_3$, Bismuth is $Bi^{+3}$, while phosphorous in $H_3PO_4$ is $P^{+5}$. In classical semiconductors such as silicon, GaAs *etc.* the HOMO (highest occupied molecular orbital) or in the language of physics, the valence band is of *p*-character and the LUMO (lowest unoccupied molecular orbital) or conduction band of *s*-character. A consequence in heavy element semiconductors, such as HgTe[50,51] or YPtBi[52], is that the lowest lying conduction band with *s* character overlaps with the highest lying valence band, see Figure 2a. However, from Hoffmann we have learnt that crossing bands are forbidden depending on their symmetry, and since spin is not a good quantum number many of the crossing points in topological materials will open up and new band gaps will appear with a band inversion (Figure 2a). The symbol for denoting normal or so-called trivial semiconductor is a donut while that for denoting inverted semiconductors is a Möbius stripe (Figure 2b). Since this effect appears in many elements it is not surprising that more than 20% of all inorganic compounds are topological.[2-4]

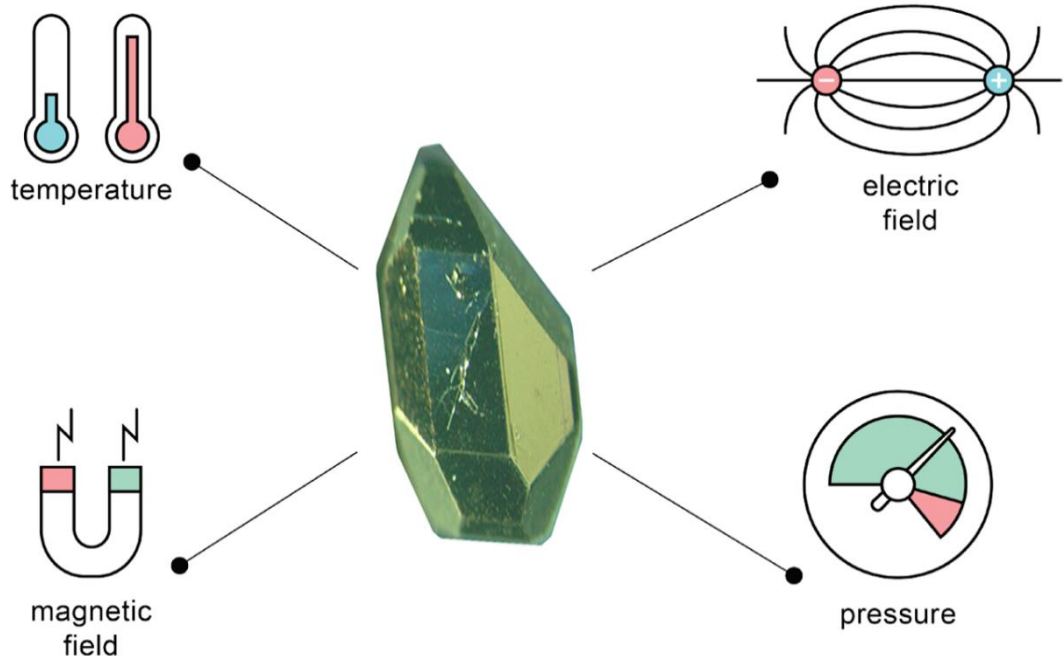

**Figure 1.** Typical external stimuli available for manipulating electronic properties of quantum materials. The single crystal at the centre of the scheme is of Weyl semimetal TaAs.

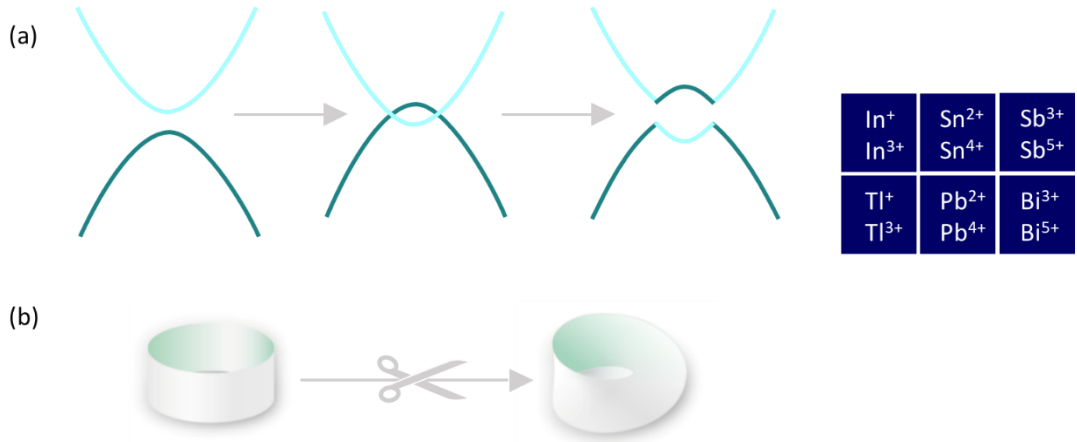

**Figure 2.** (a) Consequence of inert pair effect observed as band inversion in heavy-element semiconductors. (b) Trivial semiconductors represented as a donut while inverted semiconductor as a Möbius strip. A smooth transition is not possible between these two states.

## 2. Topological and trivial states of matter

### 2.1. Bands in insulators, metals and topological insulators

All solid materials can be broadly divided in two categories: metals, which conduct electricity and insulators, which do not. The best way to distinguish between these states is to examine their valence and conduction bands. If the valence band of a material is completely filled and separated from the conduction band by an energy gap, then it can be considered an insulator or a semiconductor (Figure 3). In an insulator such as diamond, the energy gap between the valence and conduction bands, known as the band gap, is so large that at any practical temperature, electrons cannot be excited from the valence band to the conduction band. By contrast, semiconductors such as silicon and germanium have small band gaps such that thermal excitations at high temperatures cause some electrons from the top of the valence band to populate the bottom of the conduction band. Therefore, at absolute zero temperature, because of zero thermal excitations, a semiconductor behaves as an insulator. However, in metals, the valence and conduction bands overlap with each other, and hence, the band gap is not defined. This overlap ensures that electrons are always available for conduction at absolute zero and at all finite temperatures. Furthermore, topological insulators constitute a new class of exotic materials that can neither be classified as pure insulators/semiconductors nor as metals. Inside the material or in the bulk, the valence and conduction bands are separated by the band gap, whereas on the surface, the valence and conduction bands are connected by metallic states also known as topological surface states.[53-57] Surface bands cross linearly to form graphene-like Dirac cones at the surface. One of the most important features of this surface states is that the spin and the momentum of the electrons are locked perpendicular to each other due to spin-orbit coupling. Hence, electrons carrying opposite spins propagate in opposite directions (see Figure 4b6). This prevents the backscattering of electrons at the surface of a topological insulator because it requires the electron to flip its spin. Kane and Mele presented the first prototypical example of a two-dimensional topological insulating phase in graphene.[58] Owing to the very small SOC of the carbon atom, the band gap is so small that it is not possible to access the dissipationless edge currents in graphene experimentally at any practically low temperature.[59,60] Subsequent studies focused on developing systems containing heavy elements such as $Bi_{1-x}Sb_x$ and strained α-Sn.[61,62] Groups led by Cava and Hasan provided the first experimental evidence of Dirac-like surface states on the surface of $Bi_{0.9}Sb_{0.1}$, a three dimensional topological insulator, using angle-resolved photoemission spectroscopy (ARPES).[63,64] Another approach is to go from elements to compounds. HgTe is a binary semiconductors with a heavy elements. The prediction by Bernevig *et al.*[50] for the observation of topological edge states was realized one year later by Molenkamp's group.[51] Afterwards, topological surface states were also found in layered tetradymite compounds such as $Bi_2Se_3$,[65,66] $Bi_2Te_3$,[67] $Sb_2Te_3$[68] and $Bi_2Te_2Se$[69,70] (see Figure 4). Many half-Heusler compounds having the general formula *XYZ* (*X*, *Y* are transition metals, where *X* is more electropositive than *Y* and *Z* is a main group element) are potential candidates for the realization of tunable topological insulators.[52,71,72] In fact, Liu *et al.* demonstrated the existence of topological surface states in the superconducting half-Heusler compounds, namely YPtBi and LuPtBi.[73] Owing to the excellent tunability of their structures and electronic properties through simple electron counting rules and suitability of thin film growth, Heusler compounds are critical materials for identifying new topological insulators[74]. Many more topological insulators have been discovered in inorganic solids, yet plenty of opportunities remain.[2-4]

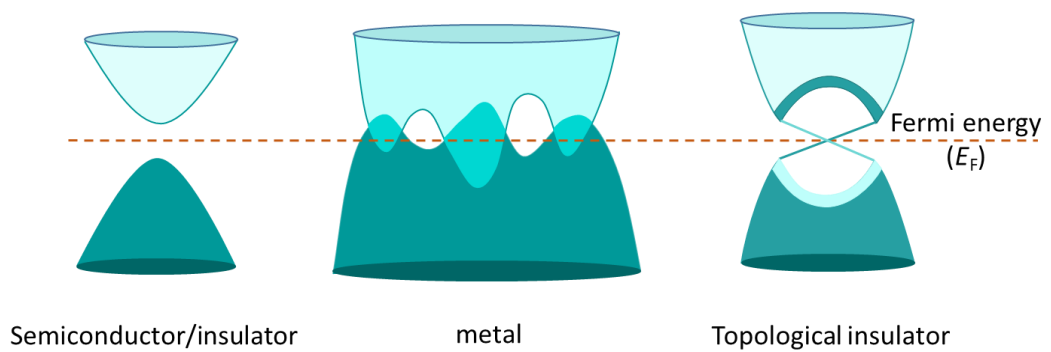

**Figure 3.** Schematics of electronic bands in various solid state materials. In semiconductors/ insulators, the valence and conduction band are separated by an energy gap. In metals, the valence and conduction bands overlap each other. In topological insulators, after band inversion, the valence and conduction bands are separated by the band gap, leaving behind a conducting topological surface states in the form of linearly crossing Dirac cone.

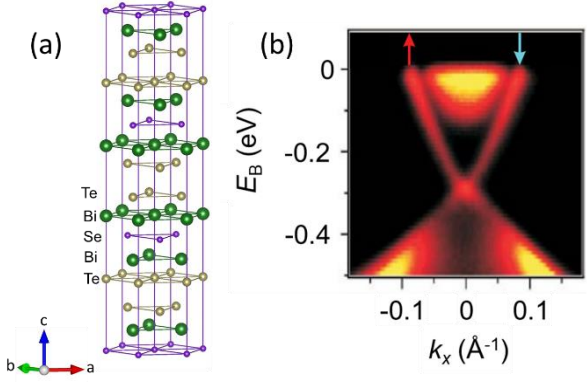

**Figure 4.** (a) Crystal structure of three dimen*m*sional topological insulator $Bi_2Te_2Se$ containing quintuple layers unit of Te-Bi-Se-Bi-Te. Te and Se atoms occupy separate layers due to their electronegativity difference. (b) ARPES results of Ca-doped $Bi_2Se_3$ showing surface Dirac cone below the Fermi energy. At the Fermi energy, the surface states co-exist with the conduction band. Upward and downward arrows depict spin-momentum locking of the surface states. Reprinted with permission from ref. [65] Copyright 2009 Springer Nature.

In three dimensional topological insulators all the crystal facets exhibit topological surface states and the number of surface Dirac cones is odd. However, some systems can support such surface states only for some facets while the remaining facets lack topological surface states. Additionally, the total number of band inversions and, therefore, the number of the surface Dirac cones is even. An even number of band crossing is often observed in more layered structures with no dispersion in one direction of the band structure. These systems are known as weak topological insulators.[53] The term "weak" was adopted because it was initially believed that the surface states were not robust against crystal disorder and an energy gap can be created in the Dirac cones by the inclusion of defects. Later, researchers found that the surface states in the weak topological insulators are quite robust against disorder, hence, its name does not fully justify these systems.[75] The best way to understand weak topological insulators is to assume a three dimensional stacking of two dimensional topological insulators with dissipationless edge states. In this case, the side surfaces will become conducting while the top and bottom surfaces, also known as the dark faces without any topological surface states will remain insulating. KHgSb is an example of the layered version of HgTe (a two dimensional topological insulator), predicted as a weak topological insulator in 2012[76], which was later identified to be a hourglass fermion[77]. The first experimental weak topological insulator phase was discovered in the hexagonal compound $Bi_{14}Rh_3I_9$ wherein the layers with conducting edges can be assumed to stack along the [001].[17] This means that the topological surface states will be absent on the surface normal to [001], while all other facets will contain even number of surface Dirac cones. Other examples of weak topological insulators are β-$Bi_4I_4$[78] and $Bi_2TeI$[79]. Interestingly, $Bi_2TeI$, in addition to containing side surface states related to weak topological insulator, also contains surface states connected to topological crystalline insulating state which we will discuss next.

In topological insulators the nontrivial metallic surface states are protected by time reversal symmetry. The topological classification of electronic structure was further extended by using knowledge of crystal structure and symmetry of the materials, which lead to the discovery of the topological crystalline insulator (TCI).[80,81] Unlike topological insulators, here the surface states are protected by the crystal symmetries, such as mirror and rotation. Till now, the TCI phase is experimentally realized in the rock salt structure type (space group $Fm\bar{3}m$) SnTe, $Pb_{1-x}Sn_xTe$ and $Pb_{1-x}Sn_xSe$.[82-84] For instance, in SnTe the topological surface states are protected by the {110} family of mirror planes and the surface states are only observed on surfaces that are perpendicular to one of the {110} mirror planes. As a consequence, the robust surface states with an even number of Dirac cones can be observed on the crystal facets such as {001}, {110} or {111}.[82,85] Since TCI surface states are protected by the crystal symmetries, it can exhibit a wide range of tunable electronic properties under various perturbations, such as, structural distortion/disorder, magnetic dopants, mechanical strain, or thickness engineering.[81]

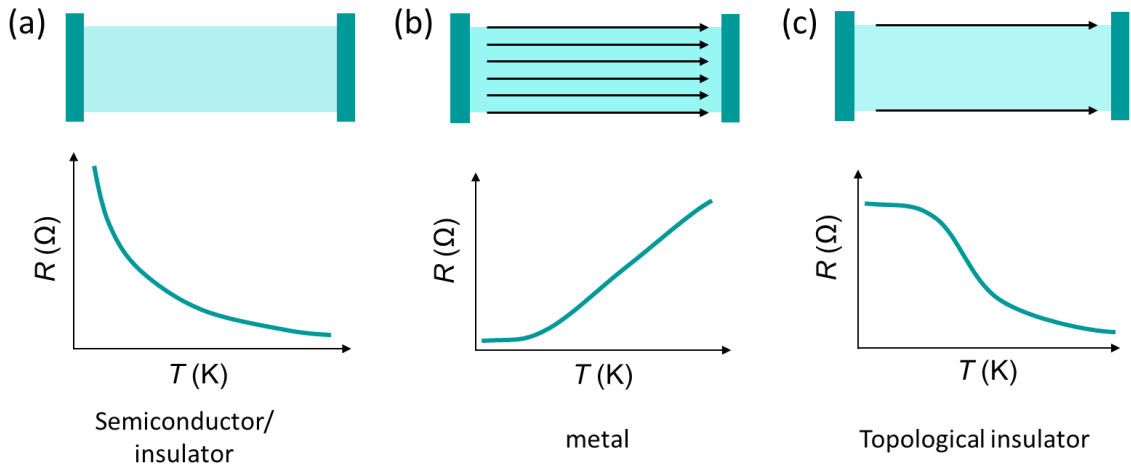

**Figure 5.** (a) Schematic representation of charge carrier conduction and typical electrical resistivity as functions of temperature in (a) semiconductors/ insulators (b) metals, and (c) topological insulators. The arrows indicate the charge carrier conduction path.

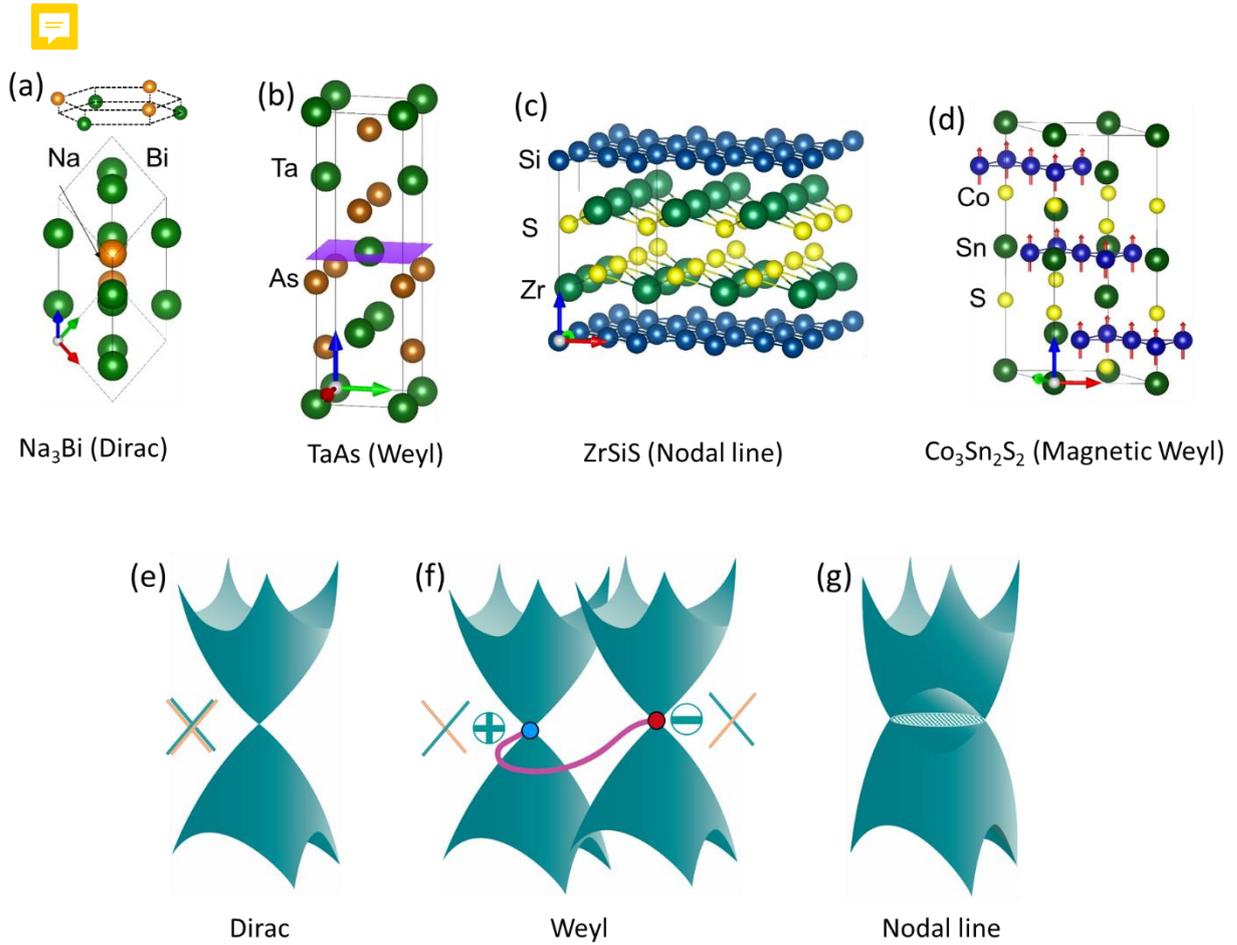

**Figure 6.** Crystal structure of (a) centrosymmetric hexagonal Dirac semimetal $Na_3Bi$, (b) noncentrosymmetric tetragonal Weyl semimetal TaAs, (c) nonsymmorphic tetragonal Dirac nodal line semimetal ZrSiS and (d) ferromagnetic kagome lattice Weyl semimetal $Co_3Sn_2S_2$. Schematic representation of band crossing in (e) Dirac, (f) Weyl, and (g) Nodal line semimetal. In a Dirac semimetal, all bands are doubly degenerate, whereas in a Weyl semimetal the degeneracy is lifted owing to breaking of the inversion symmetry or the time-reversal symmetry. Positive and negative signs indicate the opposite chirality of the Weyl points. At the surface projection, Weyl points are connected by the surface Fermi arc as depicted by the pink line. The four-fold and two-fold degenerate points are shown as crossings of 4 lines in (e) and 2 lines in (f), respectively. In case of nodal a line semimetal, the band crossing takes place in a line or a ring, where the topological surface state is two dimensional in nature, known as the drumhead surface state.

### 2.2. Transport signatures of insulators, metals and topological insulators

As the availability of electrons for conduction relies on the thermal activation of the electrons from the valence band, the resistivity of a semiconductor can be well represented using an Arrhenius equation as $\rho(T) = \rho_C \exp\left(\frac{\Delta E}{2k_B T}\right)$, where $\rho_C$ is a material constant, $k_B$ is the Boltzmann constant and $\Delta E$ is the band gap[86,87]. At high temperature, owing to the dominant thermal activation, resistivity is lower when compared to that at low temperature, where charge carriers are scarce (Figure 5a). In metals and semimetals, many free electrons are available for conduction at all temperatures; therefore, the effect of thermal excitation of electrons on conduction is much less important. Rather, conduction is affected more by temperature-dependent scattering events, the most important being the scattering of electrons by lattice vibrations (phonons). As the scattering of electrons by phonons increases at high temperature, the resistivity of the material also increases (Figure 5b). A topological insulator is the combination of a perfect surface conductor and semiconducting bulk, as reflected by its resistivity. At high temperature region, the resistivity increases as temperature decreases, similar to that observed in a semiconductor. However, at low temperature, the resistivity saturates because of the contribution from the conducting surface states (see Figure 5c). The competition between such two-channel electron transport mechanisms explains the typical temperature dependent-behavior of resistivity in topological insulators.[88-91]

### 2.3. Classification of topological semimetals

We have seen that two linear bands at the surface of a three dimensional topological insulator cross the gap between the valence and conduction bands. Therefore the question that arises is, whether such a crossing is also possible in the bulk of metals or semimetals. The answer is 'yes'. In fact, the last few years of research have been very fruitful in identifying many Dirac and Weyl semimetals, where such linear crossings of bands occur in the bulk rather than on the surface, as just discussed for the case of topological insulators. As the bulk has three spatial dimensions available, the bands around a Dirac or Weyl crossing point disperse linearly in all three momentum directions with respect to energy.[92] As both Dirac and

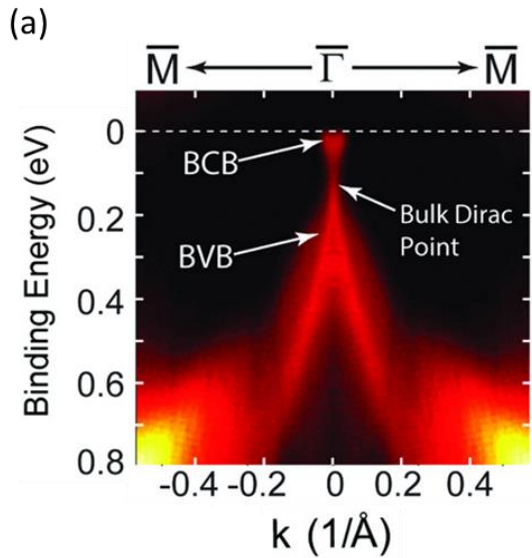
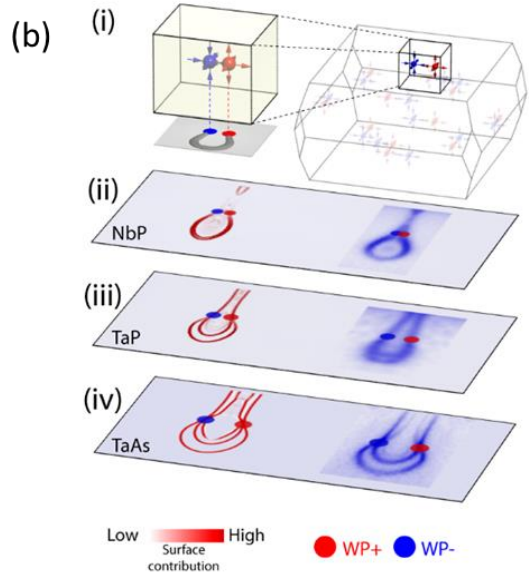

**Figure 7.** (a) ARPES results of Na$_3$Bi showing Dirac crossing between the conduction and valence band in the bulk below the Fermi energy. BCB and BVB denote bulk conduction band and bulk valence band, respectively. Reprinted with permission from ref. [95] Copyright 2014 AAAS. (b) (i) A pair of Weyl point in the bulk Brillouin zone in the TaAs family of Weyl semimetals. (ii), (iii), and (iv) show the calculated (left) and experimental surface Fermi arcs in NbP, TaP, and TaAs, respectively. The separation between the Weyl points of opposite chirality and therefore the length of the Fermi arc increases with increased SOC, i.e. the molecular weight of the compounds. Reprinted with permission from ref. [105] Copyright 2016 Springer Nature.

Weyl points result from the linear crossings of bands how do they differ from each other? Crystal structure and magnetism are the main criteria that set them apart. To realize a Weyl semimetal, one must select a system without inversion symmetry. Weyl points in centrosymmetric crystal systems cannot exist unless they also exhibit magnetism, most commonly ferromagnetism; in other words, the system should break time-reversal symmetry. In contrast, Dirac points exist in centrosymmetric crystals without magnetism (see Figure 6a). Because of the coexistence of centre of inversion in the crystal structure and the absence of magnetism, all the bands are two-fold degenerate (Kramers degeneracy) in Dirac systems, i.e. they contain both spin-up and spin-down states. Thus, the crossing of two such two-fold bands at the Dirac point would lead to four-fold degenerate states as shown in Figure 6e (represented as crossing of two pairs of lines). In Weyl semimetals, owing to the absence of a centre of inversion (see Figure 6b) or due to the presence of magnetism (see Figure 6d), the spin-up and spin-down bands are always separated, except at the high-symmetry points of the Brillouin zone (the Brillouin zone in the momentum world is equivalent to the unit cell of a crystal structure in the real world[86]). Therefore, the crossing of two non-degenerate bands would result in two-fold degenerate states at the Weyl points, as shown in Figure 6f (represented as crossing of two lines). From symmetry considerations, the minimum number of Weyl points in a nonmagnetic and magnetic Weyl semimetals are two pairs and one pair, respectively.

The most important feature of the Weyl points is that they are chiral and occur in pairs of opposite chirality. The chirality of the Weyl points is different from structural chirality; it is possible to obtain chiral Weyl points in achiral crystal structures. For simplicity, Weyl points of the opposite chirality can be considered equivalent to the north and south poles of a magnet in the real world, where the magnetic field lines originate and terminate, respectively. Similarly, the positive and negative chiralities of the Weyl points function as the source and the sink of the commonly known "Berry curvature", in the momentum world. We will discuss Berry curvature in more detail in section 5. Another important feature of Weyl semimetals is their unique unclosed arc-like topological surface states known as Fermi arcs.[93] It is defined as the connecting line between two points on the surface, the two points being the surface projections of Weyl points having opposite chiralities.

It is easy to understand Dirac points in terms of Weyl points. Dirac points are equivalent to the combination of two Weyl points with opposite chiralities; hence the net chirality of a Dirac point is zero.

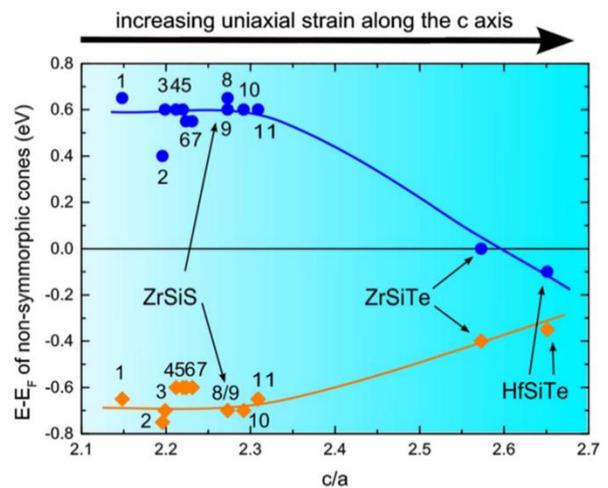

**Figure 8.** Calculated position of the nodal line with respect to the Fermi energy in ZrSiS and related compounds as a function of tetragonal c/a ratio. The nodal line in ZrSiTe resides very close to the Fermi energy. Reprinted from ref. [128] with permission under CC BY 3.0 license. Copyright 2016 Deutsche Physikalische Gesellschaft.

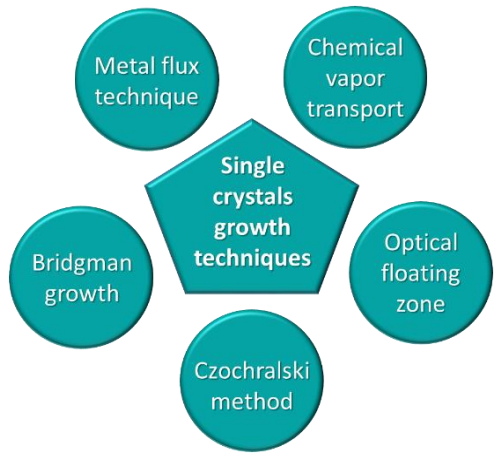

**Figure 9.** Schematic showing different single crystals growth techniques.

However, it is interesting to note that a Dirac point can be split into a pair of Weyl points with opposite chiralities by removing the center of inversion in the crystal, introducing magnetic elements into the crystal to induce ferromagnetism, or applying a high magnetic field. Three-dimensional Dirac states were first discovered in $Na_3Bi$[94,95] and $Cd_3As_2$,[96,97] where the Dirac points are protected by $C_3$ and $C_4$ rotational symmetries, respectively (see Figure 7a). From these discoveries, we can infer that the mere presence of inversion symmetry and nonmagnetic crystal structures does not guarantee Dirac points and further details of the crystal structures are vital. Furthermore, if we can remove $C_3$ rotational symmetry, as in the case of $Na_3Bi$, it is possible to eliminate Dirac points. This brings us to the next important point regarding the Weyl points, that the "Weyl crossings are accidental crossings". The term accidental means that Weyl crossings are not necessitated by crystal symmetries and are therefore completely coincident. For this reason, Weyl points are extremely difficult to eliminate, unless one finds a method to manipulate the crystal structure such that Weyl points of opposite chirality come together and annihilate each other. The first candidate compounds for realizing Weyl points were predicted in 2011 in magnetic pyrochlore irridate $Y_2Ir_2O_7$[93] and spinel $HgCr_2Se_4$[98] but could not be realized owing to various experimental difficulties. The paradigm shift in Weyl semimetal research had to wait until 2015, when simple nonmagnetic tetragonal noncentrosymmetric semimetallic compounds TaAs, NbAs, TaP and NbP were predicted to contain several pairs of Weyl points.[99] Very soon, owing to the very simple crystal growth techniques needed to grow high quality large single crystals, several research groups experimentally verified the existence of Weyl points using spectroscopic techniques in particular ARPES[100-107] (see Figure 7b) and scanning tunneling spectroscopy.[104,108] Later, Weyl points were discovered in layered transition metal dichalcogenides, namely $WTe_2$[109-112] and $MoTe_2$.[113-118] The magnetic counterparts of Weyl semimetals were identified much later in ferromagnetic Heusler compounds.[119,120] Experimental verifications of Weyl points were performed in room temperature ferromagnet $Co_2MnGa$[121,122] and ferromagnetic shandite $Co_3Sn_2S_2$.[30,31,110,123,124] These findings provide many opportunities for materials scientists to unravel many exotic electrical and thermal transport properties due to the existence of Weyl points. We discuss these properties in detail in the following sections.

In Dirac and Weyl semimetals, valence and conduction bands cross at a point. However, there can be a more general situation, where the crossing occur at a line or a ring, protected by certain crystalline symmetries. In such situations, they are called topological nodal-line or a nodal-ring semimetals (see Figure 6g). The generality of the nodal line can be highlighted by the fact that many Weyl and Dirac semimetals originate from nodal lines owing to SOC. Stabilization of such extended states requires additional crystalline symmetries, most commonly a mirror reflection. The degeneracy arising from line-crossing can be two-fold, such as that in a Weyl semimetal or four-fold, as observed in a Dirac semimetal. Nodal-ring crossing in $PbTaSe_2$,[125] a nonmagnetic superconductor, is two-fold degenerate because the crystal structure lacks inversion symmetry. Layered hexagonal structure of $PbTaSe_2$ can be described as the Pb-intercalation of $TaSe_2$, a transition metal dichalcogenide, where the Pb-layer acts as a mirror plane. This mirror plane forces the valence and conduction bands to stick together in a ring. Schoop *et al.* demonstrated that ZrSiS,[126] a Si-square-net compound (see Figure 6(c)) and related compounds[127] are also topological nodal line semimetals. As these compounds are nonmagnetic and their tetragonal crystal structures possess inversion symmetry, the nodal-line is four-fold degenerate. In such cases, nodal-lines are protected by the combination of mirror and translational symmetries (a nonsymmorphic symmetry) of the crystal structures. The position of the nodal-line in ZrSiS is much below the Fermi energy (energy levels up to which the electrons are occupied), and therefore, it is unlikely to influence the transport properties. Topp *et al.* showed that, owing to the high c/a-ratio of tetragonal structures, it is possible to move the nodal-line near the Fermi energy in ZrSiTe,[128] the sister compound of ZrSiS (see Figure 8). $Co_2MnGa$, a Heusler compound, is a ferromagnetic topological nodal-line semimetal,[122] where the nodal lines are protected by mirror planes.[32,121] Nodal lines are the origins of exotic electrical and thermal transport properties in such compounds. Nodal-ring semimetals contain flat two dimensional surface states, which are known as drumhead surface states. Drumhead surface state were first observed in topological nodal-line semimetals $PbTaSe_2$[125] and $TlTaSe_2$.[129] Later, they were also observed in the Weyl fermion line Heusler compound $Co_2MnGa$.[122] We discuss these properties in greater detail later.

## 3. Single crystals growth techniques

The ability of chemists and materials scientists to grow high-quality single crystals is one of the most important factors influencing the development of topological research. Single crystals are uninterrupted three-dimensional arrays of atoms with repeating geometry present in a single piece of a material. Growing single crystals requires considerable time and effort, as justified by their importance over their polycrystalline counterparts. Single crystals are necessary for determining the intrinsic physical properties without the influence of grain boundaries and impurity phases. They also enable observations of any anisotropic physical properties along various crystallographic axes resulting from anisotropies in the crystal structure. These advantages make single crystals particularly essential for characterizing topological materials through various transport and spectroscopic techniques. Among the many crystal growth techniques available to chemists and materials scientists, the metal-flux, chemical vapor transport (CVT), Bridgman, optical floating zone (OFZ), and Czochralski methods are popular and extremely useful to grow single crystals of a variety of topological materials (see Figure 9). The technique to be applied depends on many factors such as the volatility, thermodynamic stability, extent of doping and size of the final crystal. In the following subsections, we discuss the advantages and disadvantages of these important techniques for growing single crystals of topological materials.

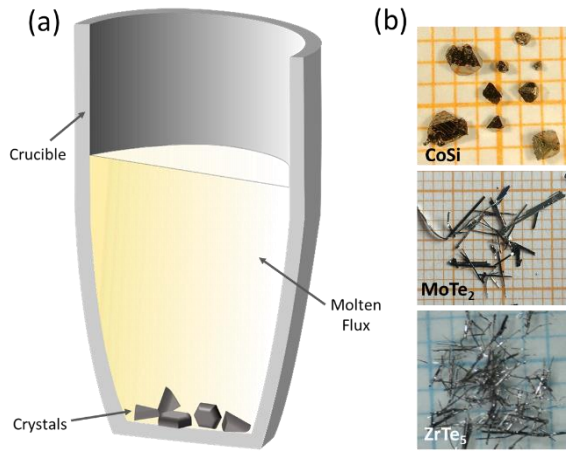
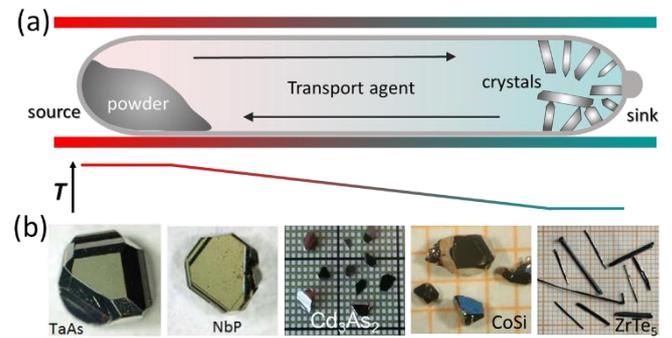

**Figure 10.** (a) Schematic of flux method. (b) Optical images of flux grown single crystals of CoSi, MoTe$_2$, and ZrTe$_5$. Panel for ZrTe$_5$ crystals: Reprinted from ref. [137] with permission under Creative Commons Attribution 4.0 International. Copyright 2018 American Physical Society.

## 3.1. Metal flux method

Sufficient diffusion of atoms for the successful growth of single crystals generally requires considerably high temperature. A metal flux that can dissolve constituent elements at relatively low temperatures can be effective to grow single crystals from the solution. Typically, when growing single crystals using the metal flux method, constituent elements are dissolved in a suitable low-melting metal flux to obtain a supersaturated solution at high temperature, which upon controlled cooling results in single crystals of the desired compositions. This method is a simple and versatile technique to grow single crystals of various compounds ranging from intermetallic Heusler compounds to relatively more ionic oxides, chalcogenides, and pnictides.[130-132] The major advantage of this method is that, it does not require specialized equipment other than a suitable crucible and a temperature-controllable furnace with homogenous temperature distribution. The most commonly used flux are bismuth, antimony, selenium, tellurium, tin, gallium, aluminium, indium, or a mixture of KCl/NaCl. Binary phase diagrams provide crucial information regarding the composition of the starting material and the temperature ranges for crystal growth.[133] The molten flux dissolves the reactant elements and forms a homogeneous solution. Once the liquid reaches the saturation limit during cooling, crystal growth is initiated. The process continues in the liquid medium up to melting point of the flux. The excess flux can be removed through simple decanting or centrifugation at high temperature. A typical setup used for the flux growth technique is shown in Figure 10a. The cooling rate is normally 1-10 K/h, but it largely depends the growth kinetics and the slope of the concentration-temperature phase. If the flux itself provides at least one of the constituents for forming the desired compound, then the technique is known as the self-flux method. Otherwise, an external flux can be selected depending upon the individual solubilities of the constituent elements. For example, LaSb crystals can be grown in a Sn flux.[134] The facets of optimally grown crystals are well developed and can easily be recognized from their shapes (See Figure 10b). Depending on the bonding anisotropies in a compound, crystals can take the forms of wires, ribbons and polygons. Flux growth is limited to those compounds that exist in equilibrium with the liquid. One of the main disadvantages of the flux method is that the metal fluxes often enter crystals as inclusions, which are then difficult to remove. Moreover, in many cases, the size of the grown crystals is not large enough for performing some bulk transport measurements, such as thermal conductivity and Nernst effect measurements.

Flux growth has been a method of choice for growing single crystals of many topological materials. Many crystals grown using the flux method can also be grown by alternative methods; however, in some cases, the flux method provides crucial advantages over other methods. One such example is Cd$_3$As$_2$, a topological Dirac semimetal. Electrical transport measurements have proved that large mobility and *MR* appear in needle-like crystals compared to polygon-shaped crystals.[25] By carefully selecting the growth temperature range during the Cd-flux growth, it is possible to drive growth along the [110]-axis to obtain needle-like crystals.[135] On the other hand, Cd$_3$As$_2$ grown using the CVT method has only polygon-shaped single crystals.[136] In some compounds where the more volatile element is used as the self-flux, it helps to maintain the stoichiometry and therefore prevent vacancies in the compound. In the Te-flux method, ZrTe$_5$ is grown in an environment having excess of Te, thereby preventing Te-vacancies as opposed to the CVT method, where the starting materials are stoichiometric amounts of elemental powders.[137] This has a profound effect on the electronic properties; the Fermi energy lies near the Dirac point in flux-grown crystals compared to CVT-grown crystals. Similarly, improved stoichiometry and low defect concentration in flux-grown crystals of Weyl semimetals, namely WTe$_2$[138] and MoTe$_2$[139] lead to excellent mobility and *MR*. Table 1 presents a comparison of these methods. Another case where the choice of the crystal growth technique is important to signify the quantum effects is the growth of CoSi, a chiral topological semimetal. Although, the crystal grown using CVT is sufficient to characterize topological features through ARPES measurements,[140,141] the transport measurements require better crystals with improved residual resistivity ratio (*RRR*) and mobility. *RRR* signifies the quality of the single crystal which we discuss in the following sections. When grown using an external Te-flux, clear quantum oscillations can be observed in magnetic field-dependent electric and thermoelectric measurements.[142,143]

## 3.2. Chemical vapor transport (CVT)

CVT has historically been used to purify solid materials in the form of single crystals. These materials can be elements, covalent compounds, ionic compounds and intermetallics. As the name suggests, the process involves transporting materials as gaseous species. Thus, all the components of the material must transform into gaseous species reversibly so that they can be redeposited as single crystals elsewhere. For the typical growth of a binary or more complex compound, the mixture of the individual elemental powders or the pre-reacted polycrystalline powder of the compound is sealed in a quartz tube under vacuum along with a transport agent. The reaction mixture is placed in a horizontal tubular furnace typically having two independent heating zones to maintain a temperature gradient (see Figure 11a). The source end of the tube containing the powder and the sink end where the crystal deposition happens are kept at two different temperatures, with a continuous temperature gradient in between providing a pathway for diffusion of the gaseous species. The deposition of the crystals occurs on the lower-temperature side or the higher-temperature side depending on whether the reaction is endothermic or exothermic, respectively. Most often, in the absence of a transport agent the vapor pressures of the reaction components are extremely small for them to be transported at the operating temperatures; therefore, a transport agent is used. The transport agent, mostly a halogen based compound (Cl$_2$, Br$_2$, I$_2$, HCl, HBr, HI and metal halides), reacts with the starting material and volatilizes it. The mass transfer of the gaseous species across the tube for redeposition as single crystals is facilitated by the applied temperature gradient. For example, consider the following reaction:[144]

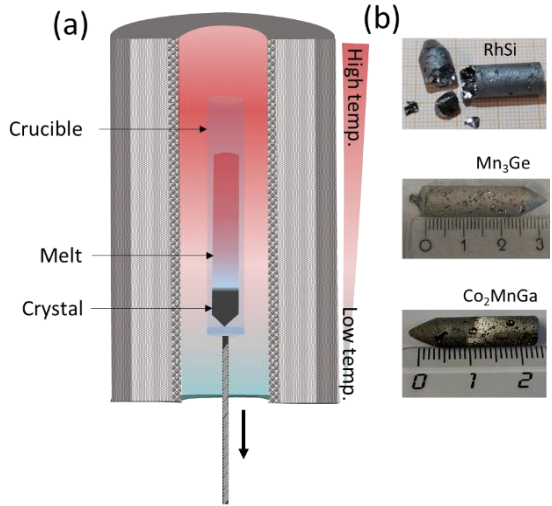

**Figure 12.** (a) Schematic of Bridgman technique (b) Optical images showing single crystals of RhSi, Mn$_3$Ge, and Co$_2$MnGa grown using this method.

$$\text{TaAs (s)} + 4\text{I}_2 \text{ (g)} \rightleftharpoons \text{TaI}_5 \text{ (g)} + \text{AsI}_3 \text{ (g)}$$

Without I$_2$ as the transport agent TaAs alone will decompose to yield solid Ta metal and As (g); hence redeposition of TaAs at the other end of the tube is not possible. However, when I$_2$ is used, it generates gaseous TaI$_5$ and AsI$_3$, which can then redeposit as TaAs single crystals. The overall thermodynamics of the reaction is particularly important. The Gibbs free energy, $\Delta G^0$, for the reaction should not be exceedingly high (typically -100 to +100 kJ/mol). If $\Delta G^0$ is highly negative, then the solid on the source side of the tube readily volatilizes; however, the redeposition as solid single crystals on the sink side would not be possible. If $\Delta G^0$ is highly positive, the solid will not vaporize and is thus not transported. Hence, an appropriate transport agent must be selected that allows for favorable reaction thermodynamics ideal temperature conditions. Because the reactions take place inside the quartz tube, the temperature cannot be exceed 1150-1200 $^\circ$C. The temperature range can be roughly estimated from the van't Hoff equation, $\ln K = \frac{-\Delta_r H^0}{RT} + \frac{\Delta_r S^0}{R}$, where, K is the equilibrium constant of the reaction; $\Delta_r H^0$ and $\Delta_r S^0$ are the enthalpy and the entropy of the reaction, respectively; and R is the ideal gas constant. For the optimum temperature, $K = 1$, and therefore, $T_{optimum} = \frac{\Delta_r H^0}{\Delta_r S^0}$.

The success story of the topological semimetals especially Weyl semimetals has largely depended on high-quality single crystals grown through CVT (see Figure 11b). Single crystals of the first experimentally verified Weyl semimetals, namely NbP, NbAs, TaP and TaAs, were grown using the CVT method.[8,102-105,145] In fact, the CVT method is the natural choice for growing single crystals of these compounds because they decompose at very high temperature before transforming into the liquid state. Hence, methods such as the Bridgman and OFZ techniques are ruled out. Moreover, using the flux method for growing large high quality phosphides and arsenides has shown limited success[146]. The quality of single crystals grown by CVT has profound role in establishing the topological properties of the materials. CVT has successfully enabled the detection of topological surface states using spectroscopic methods and has also helped prove quantum anomalies in Weyl and Dirac semimetals.[147] As mentioned in Table-1, several high-quality single crystals of topological materials have been grown using CVT, for example, WP$_2$,[148] MoP$_2$,[148] MoP,[149] WTe$_2$,[7] MoTe$_2$,[150] HfTe$_5$,[151] ZrTe$_5$,[152] and Co$_3$Sn$_2$S$_2$[30]. While many compounds listed in the table can be grown using alternative techniques, some topological materials like topological insulator β-Bi$_4$I$_4$[153,154] and Weyl Ta$_2$Se$_8$I[19,155] can only be prepared using the CVT method.

### 3.3. Bridgman method

High-temperature approaches are excellent for growing single crystals of thermodynamically stable compounds. In this context, the Bridgman method is the perfect choice for congruently melting compounds and for materials that do not undergo phase transitions between their melting point and room temperature. Congruent melting of a compound means that the chemical compositions of the solid and the melt are the same. The method is highly popular as it produces large crystals, involves fast growth, and employs relatively simple technology. Figure 12a, shows a typical Bridgman furnace. The furnace must provide a temperature gradient, which can be attained using a single heating zone or two independent heating zones. A polycrystalline compound is placed in a crucible with a sharp conical bottom. The material is heated above its melting temperature and a sharp temperature gradient is provided. The crucible is then translated slowly into the cold zone of the furnace with a constant translation velocity. When the temperature at the bottom of the crucible falls below the solidification temperature, crystal growth is initiated by the seed at the melt-seed interface. After the crucible is translated through the cold zone, the entire melt converts to a solid single-crystalline ingot. The Bridgman technique can be imployed in either a vertical (vertical Bridgman technique) or a horizontal system (horizontal Bridgman technique). The operating principles of these two configurations are similar. However, the crystals grown horizontally exhibit high crystalline quality (e.g. low dislocation density) as the crystals experience lower stress owing to the free surface at the top of the melt and are free to expand during the entire growth process. Over the years, single crystals of many topological and non-topological compounds spanning across a large variety of materials including oxides,[156,157] chalcogenides,[158-160] and intermetallics[32] have been successfully synthesized using this technique. Crystals of many topologically nontrivial materials such as RhSi,[161] Co$_2$MnGa,[32] Co$_2$VGa,[32] Mn$_2$CoGa,[32] Mn$_3$Ge,[35] and Mn$_3$Sn[162] have been synthesized using this technique (see Figure 12b).

### 3.4. Czochralski method

This growth technique is named after the Polish scientist Jan Czochralski, who first developed the method in 1916. Currently, this process is widely used for industrial production of various single crystals like Si and Ge. In the Czochralski crystal growth method, a small seed crystal is inserted from the top into the surface of a fully molten material contained in a crucible. The temperature of the melt is adjusted such that a small portion of the inserted seed is melted. Then, the seed is slowly withdrawn (usually with rotation) and a new crystal forms at the interface, as shown in Figure 13a. Generally, the new crystal grows in a cylindrical shape, whose diameter can be controlled by tuning the heating power of the melt, rotation, and pulling rate of the seed crystal. For congruently melting topological materials, the Czochralski method is an extremely powerful method to grow high-quality single crystals. Many interesting quantum materials have been grown using this technique and the most important among them are chiral compounds like PdGa[163], MnSi[164] and $Fe_{1-x}Co_xSi$[164] (see Figure 13b).

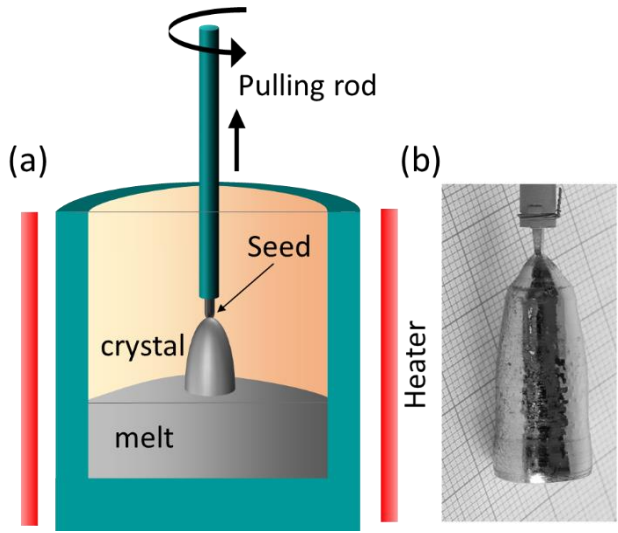

**Figure 13.** (a) Schematic of Czochralski method. (b) Optical image of a single crystal of chiral compound PdGa. Reprinted with permission from ref. [163] Copyright 2010 Elsevier Ltd.

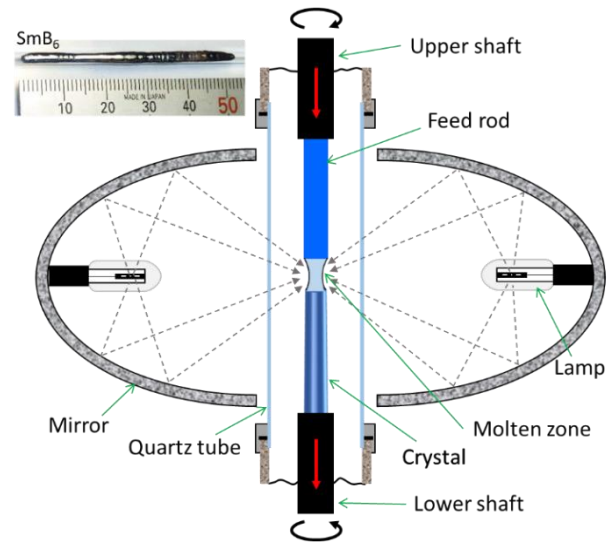

**Figure 14.** Illustration of optical floating zone (OFZ) method. Inset shows an as-grown $SmB_6$ crystal; reprinted with permission from ref. [170] Copyright 2020 Elsevier Ltd.

Although this method can be employed to grow high-quality single crystals of materials that melt congruently, it is particularly useful for chirality control when growing chiral single crystals such as topological materials CoSi, MnSi, FeSi and PdGa which crystallize in the $P2_13$ space group (structure type B20). The materials can have left- and right-handed enantiomorphs, both of which belong to the same space group. Using the Czochralski technique, one can easily control structural chirality and selectively grow left-handed or right-handed crystals. The success of this method lies in proper selection of the seed material, whose structural chirality can be transferred to the grown material. Dyadkin *et al.*[164] have shown that $Fe_{1-x}Co_xSi$ with $x = 0.08$ and $x = 0.25$ crystallize with opposite enantiomorphs. Using them as seed materials, MnSi single crystals can be grown with left- and right-handed enantiomorphs through the Czochralski technique.[164] Later, a very similar crystal growth method was used to grow opposite-enantiomorph single crystals of PdGa, a chiral topological material.[165,166]

### 3.5. Optical floating zone (OFZ)

Starting from the zone-refining technique discovered by Pfann[167] in the Bell Laboratory during the early fifties to the modern OFZ technique, this method went through a series of incremental improvements. Traditionally, this method has been used to grow single crystals of oxides. OFZ is a very useful technique to grow large single crystals of materials with extremely high melting temperature. Unlike the other methods it does not require a crucible during the growth process. The OFZ system is equipped with halogen or xenon lamp and contains two or four semi-ellipsoidal mirrors. Each mirror contains one lamp at one of its foci and the hot-zones are situated at the other foci. The four-mirror arrangement provides a more homogeneous isothermal region at the feed rod than that provided by a two-mirror arrangement. A schematic of this arrangement is shown in Figure 14. However, recently developed OFZ furnaces contain high-power laser diode units replacing the entire optical lamp and mirror arrangement. The sharp temperature profile generates a small molten zone and thus reduces material evaporation. This arrangement is particularly more effective for growing single crystals of materials containing volatile elements or the incongruent melting compounds than OFZ furnaces equipped with a halogen or xenon lamp.

In general, the light heating technique used in the OFZ furnace makes it a crucible-free method. This is a major advantage over other techniques used to grow single crystals of compounds that react with crucibles. The growth chamber in the OFZ furnace is enclosed by a quartz tube, within which various gas ambients can be maintained, such as air, nitrogen, oxygen and argon. The quartz tube also protects the heating elements from damage caused by melt-spilling and other phenomena. In general, the quartz tube can withstand gas pressures upto 10 bar, but this limit can be increased upto 300 bar using a specially designed sapphire growth chamber. The growth is usually controlled by observing the image of the molten zone captured using a charge-coupled device (CCD) camera. During the growth process, the feed and seed rods are rotated in opposite directions and the growth temperature is indirectly controlled by tuning the percentage power supplied to the lamps. The growth stability and eventually the formation of a single crystal depends on the following parameters: correct alignment of feed and seed rods, rate of crystal growth, rotation rate of mounted rods, proper tuning of growth temperature, suitable gas pressure with flow if needed, and densities of the rods. Considering the increasing interest in oxide topological materials, OFZ technique is becoming equally important for their realization in experiments. With several important and interesting materials such as carbides, borides, silicides, intermetallic Heusler topological compounds the future of OFZ in topological research appears bright. Recently, single crystals of some topological materials like $SmB_6$,[168-170] CoSi,[142] and $Co_2MnAl$[171] have been grown using this technique (see inset of Figure 14).

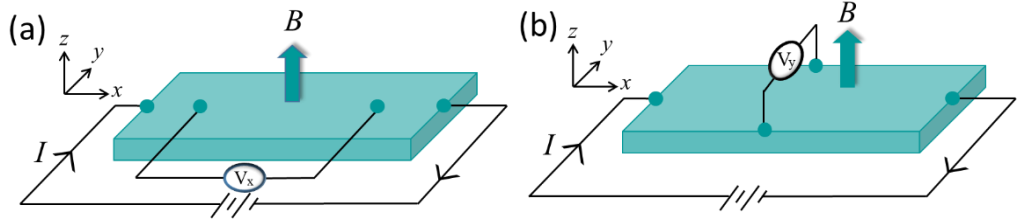

**Figure 15.** Schematic for (a) longitudinal resistivity measurements using four-probe geometry, (b) Hall resistivity measurements using four-probe geometry. The magnetic field is applied orthogonally to the applied electric current (*I*)

## 4. Electrical transport properties of topological materials

### 4.1. Electrical resistivity

Electrical resistivity is an important physical quantity indicating the extent of restriction of flow for current in a material. Using a simple four-probe geometry shown in Figure 15, by passing current from two probes and measuring the voltage drop through the other two probes, the resistance can be measured by applying Ohm's law. The voltages measured in the direction of the applied current (see Figure 15a) or perpendicular to it (see Figure 15a) give the normal or Hall resistance, respectively. This dimension-dependent resistance can be converted into dimension-independent resistivity after multiplying it with the physical dimensions of the sample being measured. The temperature, magnetic field, and pressure dependence of electrical resistivity provides various insights regarding the material characteristics. By observing the temperature dependence of resistivity, one can differentiate a metal from an insulator. For metals, resistivity increases with increasing temperature, whereas it decreases with temperature for insulators. The temperature dependence of resistivity in semimetals and semiconductors resembles that observed for metals and insulators, respectively.

Topological materials behave much like conventional materials electrically, although, with some crucial additional properties which make them special. In an ideal topological insulator, the inside or the bulk should be insulating while the surface must be an excellent conductor of electricity without any dissipation. The electrical resistivity response of a topological insulator as a function of temperature indicates that resistivity increases with decreasing temperature. This behavior is similar to that of an insulator, but instead of diverging at the lowest temperature, the curve saturates. Such a saturation in resistivity indicates conducting topological surface states. However, the challenge lies in the synthesis of an ideal topological insulator where the bulk state consists of a well-defined band gap. In real materials, owing to the underlying defects, the charge carrier concentration exceeds the limit set by the Mott criterion[172] for the system to behave as an insulator. In this scenario with metal-like resistivity, it is not straightforward to distinguish the defect-related conducting channels from the conducting topological surface states through simple four-probe resistivity measurements.

Tetradymites, with the general formula $M_2X_3$, where $M$ is a group V element such as Sb or Bi and $X$ is a group VI element such as S, Se, or Te, are the most widely studied three-dimensional topological insulators. Most of the studies on these materials have been limited to $Bi_2Se_3$, $Bi_2Te_3$, $Sb_2Te_3$, and $Bi_2Te_2Se$. $Bi_2Se_3$ is intrinsically n-type. Two of the most common defects are Se vacancies (owing to high volatility of Se) and anti-site defects, with both creating electron donating states. In $Bi_2Te_3$, Bi anti-site defects in Bi-rich reaction conditions and Te anti-site defects in Te-rich reaction conditions create acceptor and donor levels, respectively.[173] These defects in $Bi_2Se_3$ and $Bi_2Te_3$ provide several carriers in the bulk, making them behave like metals. One approach to reduce the number of carriers is to compensate the effect of donor and acceptor states by preparing a solid solution of $Bi_{2-x}Sb_xTe_{3-y}Se_y$. High semiconductor-like resistivity was observed in this solid solution owing to the compensation of acceptor states from (Bi and Sb)/Te anti-site defects and donors from Se vacancies.[174] Another elegant way is to look out for a ternary tetradymite. Here, owing to the higher electronegativity of Se than that of Te, it occupies the layers bonded to Bi from both sides. The less electronegative Te occupies the layers bonded to the other Te layer from one side and to the Bi layer from the other side (see Figure 4a). The most common defect in this compound is the Se/Te anti-site defect; because of charge neutrality, this defect does not increase the number of charge carriers.[88,91]

These carefully manipulated crystals became crucial for detailed transport investigations to analyze surface states in three-dimensional topological insulators. Peculiar oscillations in the magnetic-field-dependent resistivity at low temperature and high magnetic field ranges indicated surface-dominated transport in crystals with low charge carrier densities (~ $10^{15}$–$10^{16}$ cm$^{-3}$).[175,176] Based on the information that bulk resistivity measurements can include sizable contributions from topological surface states, researchers attempted to quantify conduction from the surface and bulk considering that the total conductance is simply the sum of the surface and bulk conductance values. Taskin *et al.* performed thickness-dependent (up to 8 μm) resistivity measurements of $Bi_{1.5}Sb_{0.5}Te_{1.7}Se_{1.3}$ using the four-probe setup and showed that surface contributions as high as 70% could be achieved at thicknesses below the thickness of 10 μm.[177] Further, it is also possible to employ a bottom-up approach such as chemical vapor deposition[178] or a top-down approach such as mechanical exfoliation to obtain few-layer-thick crystals for attaining enhanced surface transport characteristics.[179] There have also been advances in more complex multi-probe resistance measurements to separate surface and bulk conductance in topological materials.[180,181] The magnetic field dependence of resistivity in topological insulators at low temperature indicates unique, weak antilocalization[179,182] and non-zero Berry phase[204,205] in the surface states, We will discuss the concept of Berry phase in section 5.

| Table 1. Transport properties of topological materials | | | | | | | | |
|---|---|---|---|---|---|---|---|---|
| Compound | Topology type[a] | RRR (growth method) | $\rho$ (~2K) ($\Omega$ cm) | $\mu$ (2K)[b] (cm$^2$/Vs) | $n$ (~2K)[c] (cm$^{-3}$) | MR (~2K, 9T) (%) | $S_{xy}$ ($\mu$V/K) | Ref. |
| NbP | Weyl | 115 (CVT) | $6.3 \times 10^{-7}$ | $5 \times 10^6$ | $1.5 \times 10^{18}$ | $8.5 \times 10^5$ | 800 at 9 T, 109 K | 8,183 |
| TaAs | Weyl | 9 (CVT) | $5.0 \times 10^{-6}$ | $1.8 \times 10^5$ | $2.0 \times 10^{18}$ | $8.0 \times 10^4$ | 140 at 14 T, 75 K | 147,184 |
| NbAs | Weyl | 72 (CVT) | $1.0 \times 10^{-6}$ | $3.5 \times 10^5$ | $1.8 \times 10^{19}$ | $2.3 \times 10^5$ | - | 185 |
| TaP | Weyl | 11 (CVT) | $3.0 \times 10^{-6}$ | $5.0 \times 10^4$ | $2.0 \times 10^{19}$ | $1.8 \times 10^4$ | 200 at 14 T, 75 K | 145,184 |
| WTe$_2$ | Weyl | 370 (CVT) | $1.9 \times 10^{-6}$ | $4.6 \times 10^4$ | $7.1 \times 10^{19}$ | $1.7 \times 10^5$ | 31 (9T, 4 K) | 7,186 |
| | | 1256 (flux) | $4.0 \times 10^{-7}$ | $1.7 \times 10^5$ | $1.4 \times 10^{20}$ | $3.1 \times 10^6$ | | 138 |
| MoTe$_2$ | Weyl | 36 (CVT) | $4.0 \times 10^{-5}$ | $3.1 \times 10^3$ | $5.0 \times 10^{19}$ | $3.9 \times 10^3$ | - | 150 |
| | | 1064 (flux) | $9.4 \times 10^{-7}$ | $2.3 \times 10^4$ | $6.4 \times 10^{19}$ | $7.5 \times 10^4$ | | 139 |
| WP$_2$ | Weyl | 25000 (CVT) | $3.0 \times 10^{-9}$ | $4 \times 10^6$ | $5.0 \times 10^{20}$ | $4.2 \times 10^6$ | - | 148 |
| MoP$_2$ | Weyl | 2578 (CVT) | $1.0 \times 10^{-8}$ | $3.9 \times 10^5$ | $5.0 \times 10^{21}$ | $3.2 \times 10^5$ | - | 148 |
| Cd$_3$As$_2$ | Dirac | 4100 (flux) | $2.1 \times 10^{-8}$ | $8.7 \times 10^6$ | $7.4 \times 10^{18}$ | $1.3 \times 10^5$ | 80 at 1T, 300 K | 25 |
| | | 2 (CVT) | $2.0 \times 10^{-3}$ | $6.5 \times 10^4$ | $5.0 \times 10^{16}$ | $2.0 \times 10^4$ | | 187,188 |
| PtSn$_4$ | Dirac | 1025 (flux) | $3.8 \times 10^{-8}$ | $5.0 \times 10^3$ | $2.0 \times 10^{22}$ | $2.2 \times 10^5$ | 45 at 9T, 10.3 K | 189,190 |
| LaSb | Dirac | 875 (Sn-flux) | $8.0 \times 10^{-8}$ | $4.4 \times 10^5$ | $1.1 \times 10^{20}$ | $9.0 \times 10^5$ | - | 134 |
| LaBi | Dirac | 339 (flux) | $1.5 \times 10^{-7}$ | $1.8 \times 10^4$ | $1.5 \times 10^{21}$ | $8.2 \times 10^4$ | - | 191 |
| PtBi$_2$ | Dirac | 1667 (flux) | $2.4 \times 10^{-8}$ | $5.5 \times 10^4$ | $2.0 \times 10^{20}$ | $1.4 \times 10^6$ | - | 192 |
| ZrTe$_5$ | Dirac | - (CVT) | $1.2 \times 10^{-4}$ | $2.7 \times 10^4$ | $1.9 \times 10^{17}$ | - | - | 193 |
| | | - (flux) | $1.1 \times 10^{-4}$ | $5.0 \times 10^5$ | $1.5 \times 10^{17}$ | $8.2 \times 10^3$ | 200 at 8 K | 194,195 |
| HfTe$_5$ | Dirac | - (CVT) | $1.9 \times 10^{-3}$ | $3.8 \times 10^3$ | $1.3 \times 10^{18}$ | $1.5 \times 10^3$ | 600 at 100 K, 4T | 196,197 |
| | | - (flux) | - | $2.8 \times 10^4$ | $8.9 \times 10^{16}$ | $9.0 \times 10^3$ | - | 198 |
| PbTaSe$_2$ | Nodal line | 115 (CVT) | $2.8 \times 10^{-7}$ at 4K | - | - | - | | 199 |
| ZrSiS | Nodal line | 300 (CVT) | $4.8 \times 10^{-8}$ | $6.3 \times 10^3$ | $2.1 \times 10^{22}$ | $6.0 \times 10^4$ | | 200 |
| HfSiS | Nodal line | 10 (CVT) | $3.1 \times 10^{-6}$ | $2.4 \times 10^3$ | $4.5 \times 10^{20}$ | $5.8 \times 10^2$ | | 201 |
| NbAs$_2$ | Nodal line | 317 (CVT) | $2.8 \times 10^{-7}$ | $1.6 \times 10^5$ | $3.1 \times 10^{19}$ | $8.8 \times 10^3$ | | 202 |
| TaAs$_2$ | Nodal line | 100 (CVT) | $1.3 \times 10^{-6}$ | $2.2 \times 10^3$ | $2.4 \times 10^{19}$ | $2.0 \times 10^5$ | - | 203 |

[a] without magnetic field, [b] $\mu = \sqrt{\mu_e \mu_h}$, [c] $n = n_e + n_h$

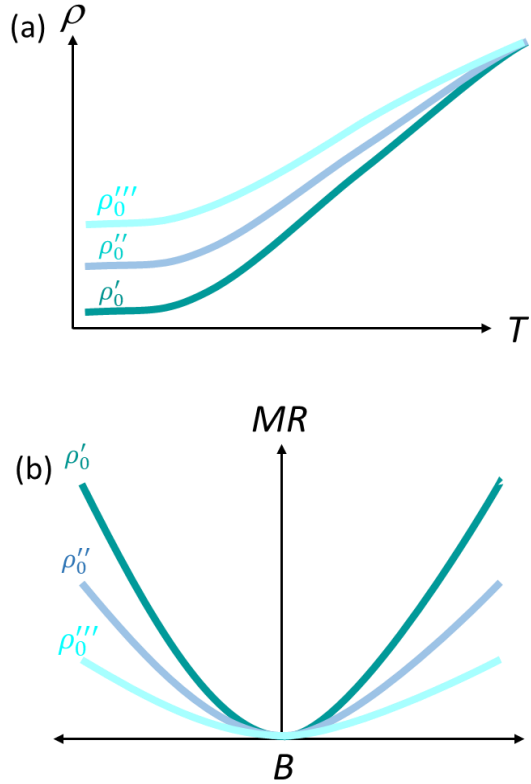

**Figure 16.** (a) Schematic of temperature-dependent resistivity for different crystal qualities. The higher the crystal quality, the larger the residual resistivity ratio (*RRR*). (b) Magnetic field dependence of resistivity response curve for different crystal qualities. Here, $\rho_0'$, $\rho_0''$ and $\rho_0'''$ denote the residual resistivity for crystals of different purities. The lower the residual resistivity value, the higher is the crystal purity and *MR*.

Majority of Weyl and Dirac materials are semimetals and their room-temperature resistivity values are rather high compared to those of metals. Resistivity in semimetals/metals can be simply understood in terms of the scattering of mobile electrons by the most common scatterers such as defects, phonons, or other electrons. In the free- electron picture, electron–electron scattering can be neglected. Electron–phonon scattering dominates at high temperature, whereas defect scattering is dominant only at low temperature and has no temperature dependence. Resistivity in a semimetal continuously decreases with decreasing temperature, but it changes very little at low temperature. This remaining resistivity at the lowest temperature (commonly 4.2 K or lower, i.e. liquid-He boiling temperature) is called residual resistivity ($\rho_0$). Because defects are the most dominant scatterers of electrons at such low temperatures, the value of $\rho_0$ is governed by defect concentration in the sample. Hence, to compare the cleanliness of various single crystals of the same compound, it is customary to compare their residual resistivity ratios, *RRR*, which is the ratio of the resistivity at room temperature and residual resistivity $\rho_0$ $\left(RRR = \frac{\rho_{300K}}{\rho_0}\right)$. The crystal with a smaller defect concentration, i.e. smaller $\rho_0$ or higher purity, shows higher *RRR* and vice versa, as indicated in Figure 16a. Measuring temperature- dependent resistivity to obtain *RRR* is a very effective tool to screen pure crystals for detailed investigations without complex and time-consuming microscopic characterization. A pure crystal with low $\rho_0$ is the key to attain an enhanced mean free path, mobility, and *MR* in semimetals. Clear evidence of $\rho_0$-dependent mobility and *MR* in topological semimetals such as $Cd_3As_2$,[25] a Dirac semimetal, and $WP_2$[148] a Weyl semimetal, has been provided. A schematic depicting the effect of *RRR* on *MR* is shown in Figure 16b.

Till now, we have considered normal resistivity without a magnetic field. A magnetic field is an effective tool to manipulate the motion of electrons in metals; however, the effects observed in topological semimetals are much more dramatic compared to those in other materials. The most common effect is the change in resistance on applying a magnetic field perpendicular to the applied current at a constant temperature, known as *MR*. It is calculated as $= \frac{\rho(B)-\rho(0)}{\rho(0)}$, where $\rho(B)$ and $\rho(0)$ are the resistivity values in the presence and absence of a magnetic field, respectively. Generally, Dirac and Weyl semimetals exhibit extremely high *MR* at low temperature owing to the combination of two main factors: (a) The availability of pure samples ensures that zero-field resistivity $\rho(0)$ below the liquid-He boiling temperature ($\rho_0$) is small, which automatically enhances *MR* according to the above equation. (b) Special topological features in the band structure such as linearly dispersed bands near the Dirac or Weyl points ensure high mobility at low temperature. Mobility is discussed in the next section. Mobility in conventional metals is generally low compared to that in semimetals; however, owing to the presence of a large number of carriers, conductivity is high. On the other hand, mobility in semimetals, especially topological semimetals, is very high; however, owing to the small number of charge carriers, conductivity is not as high as that in noble metals. Figure 17c shows some typical topological semimetals and metals arranged according to their *MR* and conductivity at 2 K. Ideal metals such as K and Cu behave as materials with high conductivity and low *MR* and the topological semimetals, except for a few, behave as materials with low conductivity and high *MR*. Few topological semimetals such as $WP_2$, $MoP_2$, and $Cd_3As_2$ exhibit both high *MR* and conductivity together. A huge suppression of scattering of electrons at high angles or backscattering from defect sites have been reported in the latter compounds, imparting high conductivity at low temperatures.[25,148] In fact, fluid-like flow of electrons or the hydrodynamic effect along with strong violation of Wiedemann-Franz law has been observed in Weyl semimetal $WP_2$.[26]

One of the main features of *MR* in topological semimetals apart from its typically high value is its parabolic nature; this shape is retained up to very high magnetic field (see Figure 17b). This parabolic *MR* indicates carrier compensation in the system, implying that the numbers of electron and hole carriers are equal.[7] Considering that the mobility of electrons and holes are equal ($\mu$), then the resistivity as a function of magnetic field can be expressed as:

$$\rho(B) = \frac{(n+p)}{e\mu} \frac{1+(\mu B)^2}{(n+p)^2 + (n-p)^2(\mu B)^2}$$

where, *e* is the electronic charge and *n* and *p* are number of electrons and holes, respectively.[206,207] For compensation condition ($n = p$), the above equation reduces to $MR = (\mu B)^2$, i.e. a parabolic *MR*. Consider an extreme case of uncompensation, where, $n \gg p$. In this scenario, $(n+p) \approx (n-p) \approx n$. Hence, the resistivity does not vary with magnetic field and therefore saturates. Owing to the domination of a single type of charge carrier, saturation of *MR* under moderate fields is observed in metals.

In the presence of a magnetic field, the electrons inside a metal can be imagined as performing circular or cyclotron motions. Under a high magnetic field, the product of the cyclotron frequency, $\omega_c$, and the scattering time, $\tau$, (the average time between two scattering events) is much greater than unity, i.e. $\omega_c\tau \gg 1$. In this scenario, the continuous occupied energy states in the band become discrete (Landau levels) and the energy gap between two Landau levels scales with the magnetic field. This manifests an oscillatory nature of resistivity if measured as a function of the applied magnetic field, known as the Shubnikov–de Haas (SdH) oscillation. Apparently, this oscillation can also be observed in several other physical quantities such as magnetization (de Haas-van Alphen oscillations, dHvA), thermopower and specific heat.[208] At high temperature, the separation between two Landau levels becomes lower than the thermal energy, i.e. $\hbar\omega_c < k_BT$, and hence, the levels broaden and cease to exhibit oscillations. Therefore, quantum oscillation studies must be conducted at sufficiently low temperature. Quantum oscillation is an excellent tool to study the shapes and sizes of Fermi surfaces. However, quantum oscillations can also be used to differentiate between a topological and a non-topological band,[209] to locate the Dirac or Weyl points with respect to the Fermi energy level and calculate the effective masses of the charge carriers. Furthermore, in addition to using a high magnetic field, one must select a pure single crystal such that $\omega_c\tau \gg 1$. In a dirty sample, the electrons are scattered much before completing a full cyclotron motion, thereby destroying the oscillatory signal. It should be notes that quantum oscillation technique, especially dHvA oscillation is a bulk sensitive technique, therefore are not suitable to study the surface states. However, Qu *et al.* observed clear SdH oscillations from the surface states of topological insulator Bi$_2$Te$_3$, because owing to the insulating bulk, all the current is carried only by the conducting surface.[175]

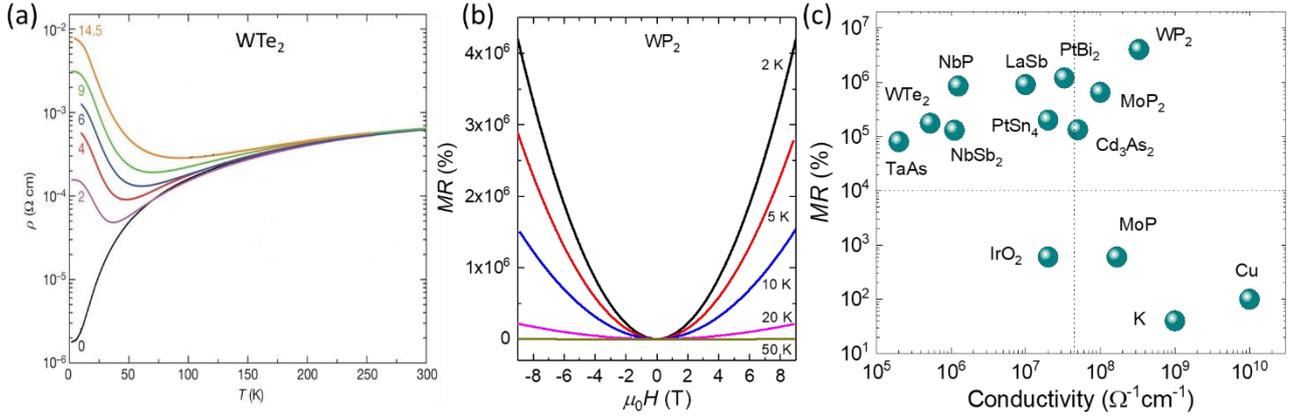

**Figure 17.** (a) Temperature dependence of resistivity in WTe$_2$ measured under different magnetic fields. Reprinted with permission from ref. [7] Copyright 2014 Springer Nature. (b) Magnetoresistance (*MR*) data for WP$_2$ crystal under magnetic field up to 9 T at different measurement temperatures. (c) Electrical conductivity vs. MR plot for different topological semimetals and some ideal trivial metals. Panel (b) and (c): Adapted from ref. [148] with permission under CC BY 4.0 license. Copyright Springer Nature.

Weyl and Dirac semimetals are also known to exhibit exotic transport phenomena like chiral anomaly induced negative *MR*[147,210] and planar Hall effect[211,212] due to their topological band structure. In a Weyl semimetal, the number of fermions of opposite chiralities is equal in the absence of electromagnetic field, thus maintaining a chiral symmetry. However, in the presence of parallel electric field (current) and magnetic field this chiral symmetry is broken by a net pumping of chiral fermions between Weyl points of opposite chiralities. This phenomenon is known as chiral anomaly, which in the transport measurements can be observed as positive magnetoconductance or negative *MR* on the application of electrical current and magnetic field parallel to each other.[213] This effect has also been observed in Dirac semimetals, because a Dirac point splits into a pair of Weyl points on applying a magnetic field.[214]

### 4.2. Electron hydrodynamics

Hydrodynamical flow is characterized by the collective flow of particles, for example, the flow of water in a pipe. Owing to the viscosity of the fluid, the layer adjacent to the wall of the pipe moves slowly compared to the layer at the centre. The picture of the flow of electrons in a metallic wire is quite different. Electron transport in metals can often be explained qualitatively by the Drude model. It assumes that on the application of an electric field the electrons drift independently and slow down only when encountered by impurities or phonons that provide electrical resistance. However, Gurzhi proposed that if the electrons in a material interact much more often among themselves compared to interacting with impurities or phonons, then a collective hydrodynamic-flow of electrons can be achieved.[215] In this scenario a shear viscosity of the electron fluid can be defined such that the electrons at the centre will move faster compared to those at the boundary. The resistivity, which does not depend on the dimensions of a sample in normal electrical transport, will now increase on decreasing the cross sectional area of the sample. In most materials, the interaction of electrons with impurities at low temperatures and the interaction of electrons with phonons at high temperatures are so strong that hydrodynamic flow is not observed. Since the electron-phonon interaction at high temperatures cannot be avoided, one must look for very high quality single crystals of materials with very small impurity concentrations. Researchers have considered two dimensional electron gases in (Al,Ga)As heterostructures[216], graphene[217,218] and delafossite PdCoO$_2$[219] for the observation of the hydrodynamic effect of electrons at low temperatures. Sulpizio et al. used a scanning carbon nanotube single-electron transistor to image the behaviour of electron transport locally in a high mobility graphene sample.[220] Interestingly, the current density and the Hall voltage profile across the width of the sample both mimic the velocity profile of flow of a liquid in a pipe.

Dirac and Weyl semimetals exhibiting large *RRRs* and mean free paths are a natural choice for studying hydrodynamic effects. Gooth *et al.* performed width-dependent transport measurements of the Weyl semimetal WP$_2$ and demonstrated that the resistivity depended inversely on the square of the width, a clear indication of the hydrodynamic effect.[26] The hydrodynamic effect was associated with a strong violation of the Wiedemann-Franz law, a law which states that the ratio of the electronic thermal conductivity and electrical conductivity is a universal constant for metals at a given temperature. Subsequently, there have been attempts to understand the mechanism of hydrodynamic effects in Weyl semimetals. Coulter *et al*. show that the hydrodynamic effect in WP$_2$ is not mediated by momentum conserving electron-electron scattering because the associated scattering time is much longer than other scattering processes which relax the momentum of electrons.[221] Rather, it was found that momentum conserving scattering process essential for the hydrodynamic effect is phonon mediated. In fact, it was predicted that the majority of the electron-phonon scattering at low temperatures (scattering of electrons by acoustic phonons) alters the path of the electrons by only small angles and, therefore, the momentum remains quasi-conserved. Additionally, there are some indications of phonon-mediated electron-electron scattering which can conserve momentum. In Weyl semimetal WTe$_2$, phonon-mediated electron-electron scattering becomes most dominant above 10 K among all other possible scattering phenomena.[222] In order to verify a hydrodynamic electron flow in WTe$_2$, Vool *et al*. used the nitrogen-vacancy centre to detect the local magnetic field that originates from the current flow in a micro-crystal.[222] The observation of larger current density at the centre of the specimen compared to that at the boundary above 10 K confirms the hydrodynamic current flow in WTe$_2$. Some recent Raman spectroscopic studies further signify the importance of electron-phonon scattering among microscopic scattering processes on the transport properties of Weyl semimetals.[223,224] Since, electron-electron interaction is not a dominating factor for most of the topological semimetals, understanding the electron-phonon interaction is essential to explore hydrodynamic effects in this broad class of materials.

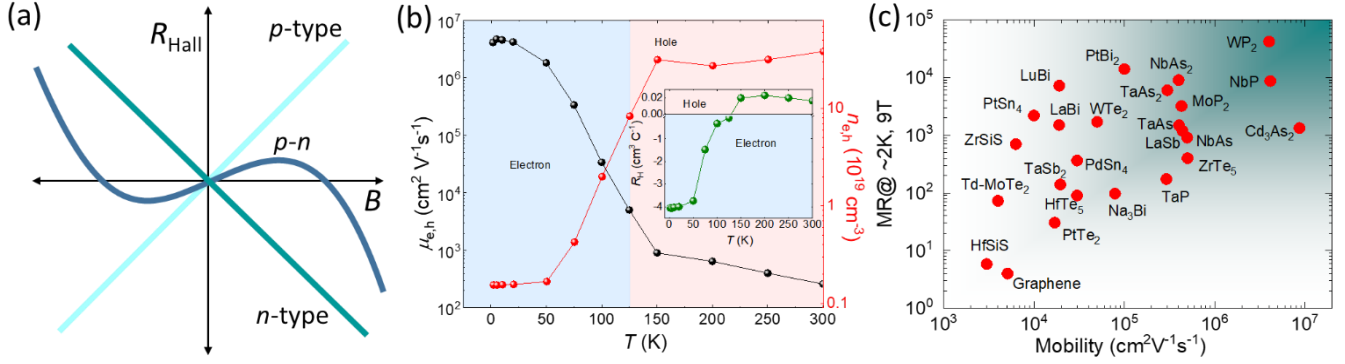

**Figure 18.** (a) Illustration of the Hall response curve as a function of the magnetic field for different dominant carrier type (*p*, *n* and *p-n*) material. (b) Temperature dependence of the mobility (left ordinate) and the carrier density (right ordinate). Reprinted with permission from ref. [8] Copyright 2015 Springer Nature. The inset shows the evolution of the Hall coefficient with temperature. The temperature regimes where electrons and holes act as the main charge carriers are indicated by the blue and red shading, respectively. (c) Charge-carrier mobility vs. *MR* plot for different topological crystals at 2 K and 9 T magnetic field. Reprinted from ref. [230] with permission under CC BY 3.0 license Copyright 2016 IOP Publishing.

### 4.3. Hall effect

The Hall effect was first discovered by Edwin Hall in 1879.[225] It is simply the manifestation of the Lorentz force experienced by electrons or charge carriers in semiconductors and metals in the presence of a magnetic field applied perpendicular to the current, as shown in Figure 15b. The Lorentz force is perpendicular to both the carrier velocity (electric current) and magnetic field. The charge carrier accumulation at the edge of the sample continues until it is eventually ceased by the electric field experienced in the opposite direction, which is considered to be the equilibrium state. Hall voltage ($V_y$) is a measure of the voltage across the sample edges in the *y*-direction, at the equilibrium state. Hall resistance ($R_{yx}$) can then be expressed as $R_{yx} = \frac{V_y}{I_x}$, where $I_x$ is the current passed along the *x*-axis. The Hall effect is one of the most fundamental phenomena for understanding the electronic properties of metals and semiconductors. The Hall constant ($R_H$) is the electric field generated along the *y*-axis when a unit magnetic field and a unit electric current are passed along the *z*- and x-axes, respectively. $R_H$ is an intrinsic quantity specific to a particular material. The relation $R_H = \frac{\Delta \rho_{yx}}{\Delta B}$, where $\Delta \rho_{yx}$ is the change in Hall resistivity and $\Delta B$ is the change in magnetic field, is used to calculate the experimental $R_H$. The sign of $R_H$ provides essential information on whether the majority charge carrier is electron-type (negative $R_H$) or hole-type (positive $R_H$). Various expressions of $R_H$ can then be used to calculate the important parameters of the materials' electronic properties. The relation $R_H = \frac{1}{ne}$ provides an estimation of the charge carrier density ($n$), where *e* is the charge of electron. As $R_H$ is inversely proportional to the charge carrier density, it increases in the order semiconductor→semimetal→metal.

Mobility ($\mu$) refers to the speed at which the carriers traverse in a material when a unit electric field is applied in a material. This parameter can be calculated from the relation $\sigma = \mu n e$ if the conductivity ($\sigma$) in the absence of a magnetic field is known. The mobility of charge carriers is considerably higher in a single crystal than in polycrystalline materials because carriers are scattered off the grain boundaries in polycrystals, thus reducing the average velocity of the carriers. The mobility in a single crystal can be enhanced by increasing the purity or decreasing the defect concentrations, particularly at low temperatures where the defect-related scattering of charge carriers play a more important role. For this reason, the mobility of semimetals or metals increases systematically with an increase in the *RRR* value. Historically, mobility has been an important electronic parameter in semiconductors, for example in transistors and other devices. However, the information that mobility provides on semimetals is also extremely crucial. The mean free path, which is the average distance an electron or hole travels before it is scattered, is directly proportional to the mobility. Large mean free paths or simply, the accessibility of the highest-quality single crystals is essential for observing the quantum phenomena in semimetals and metals. An example of these phenomena is quantum oscillation in physical quantities such as magnetization, resistivity, and specific heat.[208] Graphene is particularly known for its large mobility, which exhibits a value of $1.5 \times 10^4$ cm$^2$ V$^{-1}$ s$^{-1}$ at room temperature[226] and has the potential to reach as high as $1 \times 10^5$ cm$^2$ V$^{-1}$ s$^{-1}$ upon sample improvement.[227] The large mobility is one of the main reasons why graphene exhibits the quantum Hall effect at room temperature.[228] Among the bulk semimetals, bismuth warrants special mention because it holds the record for the largest mean free path of the order of a millimeter at low temperature[229]. Recent advancement in topological Dirac and Weyl semimetals has facilitated the exploration of numerous materials that demonstrate extremely large mobilities. Liang *et al.* studied a number of crystals of Dirac semimetal $Cd_3As_2$ and revealed the evolution of mobility as a function of crystal quality.[25] The crystal with the largest low temperature conductivity or the largest *RRR* exhibits the maximum mobility and vice versa. The crystal with the largest low-temperature conductivity or the largest *RRR* exhibits the maximum mobility and vice versa. The highest-quality crystal (*RRR* = 4100) exhibits extremely large mobility ($9 \times 10^6$ cm$^2$ V$^{-1}$ s$^{-1}$) at 5 K. Presumably, a strong protection of electrons from backscattering (reversing the momentum of electrons) is the effective prolongation of the mean free path and hence, a large mobility. The Weyl semimetal NbP with a comparatively small *RRR* of 115 exhibited a mobility of $9 \times 10^6$ cm$^2$ V$^{-1}$ s$^{-1}$ at 1.85 K[8] (see Figure 18b), which is the same order of magnitude as that observed in $Cd_3As_2$. This means that even a considerably higher mobility is possible if one finds a way to enhance the crystal quality and increase *RRR*.

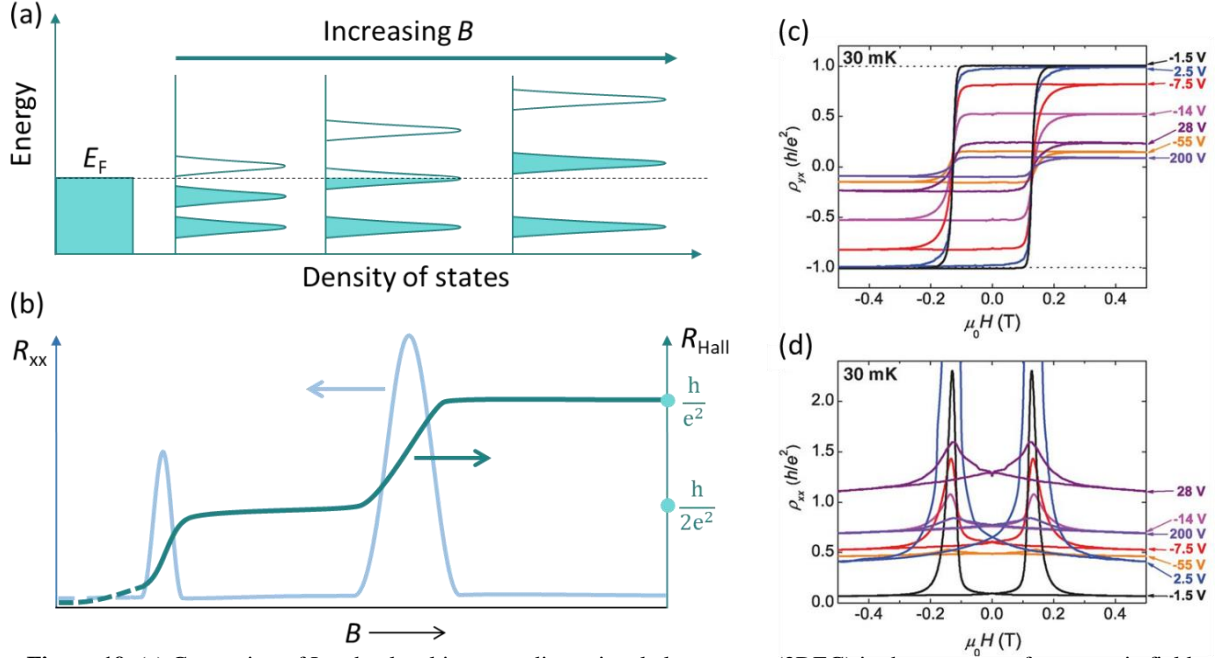

**Figure 19.** (a) Generation of Landau level in a two-dimensional electron gas (2DEG) in the presence of a magnetic field. On increasing magnetic field the separation between the Landau levels increases and after a critical field only the last Landau level remains below the Fermi energy, representing the extreme quantum limit. (b) Corresponding Hall and longitudinal resistance with respect to the magnetic field. The Hall resistance exhibits plateaus and the corresponding longitudinal resistance is zero except at fields where the transition occurs between two plateaus. (c) Hall resistivity of the thin film of Cr-doped $(Bi,Sb)_2Te_3$ at various gate voltages at 30 mK. Complete quantization ($\frac{h}{e^2}$) at the gate voltage of -1.5 V is observed. The corresponding zero value of longitudinal resistivity in (d) signifies the dissipationless edge current. Reprinted with permission from ref. [234] Copyright 2019 AAAS.

As previously discussed, if a particular type of charge carrier dominates in a material, then the data of the Hall resistivity as a function of the magnetic field is linear, with a negative or positive slope for *n*- or *p*-type materials, respectively, as shown in Figure 18a. This is typically the case for semiconductors and metals. However, multiple valence or conduction bands can cross the Fermi energy of semimetals. Therefore, both the electron and hole charge carriers of comparable amounts can exist simultaneously. In such cases, the behavior of the Hall resistivity vs the magnetic field is no longer linear and can even exhibit *n-p* transition at an intermediate value of the magnetic field, as shown in blue in Figure 18a. At this point, the simple methodology of using $R_H = \frac{\Delta \rho_{yx}}{\Delta B}$ to calculate the Hall constant is no longer applicable. The estimation of the concentration and mobility of individual carrier types is not very simple. However, the working formula expressed as equation:

$$\sigma_{xy} = \left[ \frac{n_e \mu_e^2}{1 + (\mu_e B)^2} - \frac{n_h \mu_h^2}{1 + (\mu_h B)^2} \right] eB$$

is most commonly used. The magnetic field-dependent Hall conductivity ($\sigma_{xy}$) must be fit using $n_e$ (electron concentration), $n_h$ (hole concentration), $\mu_e$ (electron mobility), and $\mu_h$ (hole mobility) as the fitting parameters. This method is only an approximation because of the number of fitting parameters involved. The mobility of many important topological materials are outlined in Table 1 and summarized in Figure 18c. Figure. 18c depicts the arrangement of *MR* as a function of mobility. This figure shows a general trend that the *MR* increases with mobility, which is quasi-independent of the material being investigated.[230] This reflects the point that electrons (or holes) are generally protected against backscattering in topological materials, thus providing the range of mobility observed. The extent of protection will depend on the details of the topological band structure of a particular compound and the crystal quality. Once the protection against backscattering is removed upon the application of the magnetic field, the resistance will be maximum for the largest mobility, thus explaining the increasing trend of *MR* with mobility.

### 4.4. Quantization of the Hall effect

From the viewpoint of condensed matter physics, the field of topology is generally regarded as having started almost exactly 40 years ago with von Klitzing's discovery of the QHE in a two-dimensional electron gas (2DEG).[231] The 2DEG was formed in a complex heterostructure, where the movement of the electrons are restricted to a two-dimensional plane. The position of the Fermi energy could be tuned by gating with an electric field to enable the QHE to be observed at extremely low temperatures in high magnetic fields.[231] Alternatively, the same effect is observed at a fixed gate voltage when the magnetic field is varied. The Hall resistance does not vary continuously with the magnetic field (or gating voltage), but instead exhibits steps (plateaus) (see Figure 19b). The values of the conductance at the plateaus are quantized to integer multiples of e$^2$/h, i.e., $G_{xy}(B) = v\frac{e^2}{h}$, where e is the charge of the electron, h is Planck's constant and $v$ is an integer that is also known as the filling factor. The longitudinal resistance remains zero, except at the fields where the transition occurs from one plateau to another. This zero resistance indicates dissipationless electrical current at the edge of the device. The entire surface, except the edges, behaves as an insulator because the Fermi energy lies in the gap between two successive Landau levels. This insulating behavior differs from that of a trivial insulator and hence requires topological effects to understand it. As a materials scientist one dreams immediately of finding a material in which the QHE could be observed at room temperature and in the absence of a magnetic field. One dream became true in graphene, an innately two-dimensional material in which the QHE was observed at room temperature. Nowadays, - almost every day it seems! - new exciting physics related to graphene appears.[232] In 1988, Duncan Haldane developed a model based on a honeycomb lattice, which allowed for the QHE without a magnetic field in magnetic materials, i.e., the quantum anomalous Hall effect (QAHE).[233] This proposal was not experimentally feasible; however, we will discuss the QAHE in the next paragraph. Meanwhile, Haldane's work inspired several other theoreticians, including Kane and Mele, to propose the quantum spin Hall effect (QSHE) in graphene,[58] where spin-orbit coupling should do the work of the intrinsic magnetic field. Under this setting, the electrons at the edges move without any dissipation like in the QHE, but the SOC forces the up-spin and down-spin electrons to move in opposite directions. However, the SOC effect in graphene is too small, thus requiring other elements or compounds formed with heavy elements. Bernevig, Hughes, and Zhang, recognized that it is possible to obtain an inverted band gap in the quantum well structure of mercury telluride–cadmium telluride semiconductors by increasing the thickness of the mercury telluride layer beyond a critical thickness.[50] In this scenario, the dissipationless edge current is possible, even without the application of a magnetic field. Molenkamp's team immediately recognized the potential impact of binary semiconductors with heavy elements. The experimental realization and the first observation of the QSHE was achieved less than a year after its prediction by demonstrating the quantized conductance of $\frac{2e^2}{h}$.[51] The search subsequently began for three-dimensional topological insulators where the entire surface, rather than the edge, can support dissipationless current. This rapidly resulted in the substantial research area of topological materials science. Unfortunately, no single element of the periodic table fulfils the conditions required for a three-dimensional topological insulator or a two-dimensional QSHE. However, the Princeton teams led by Cava and Hasan recognized that the Bi$_2$Se$_3$ family was an excellent candidate material for three-dimensional topological insulators.[65,66]

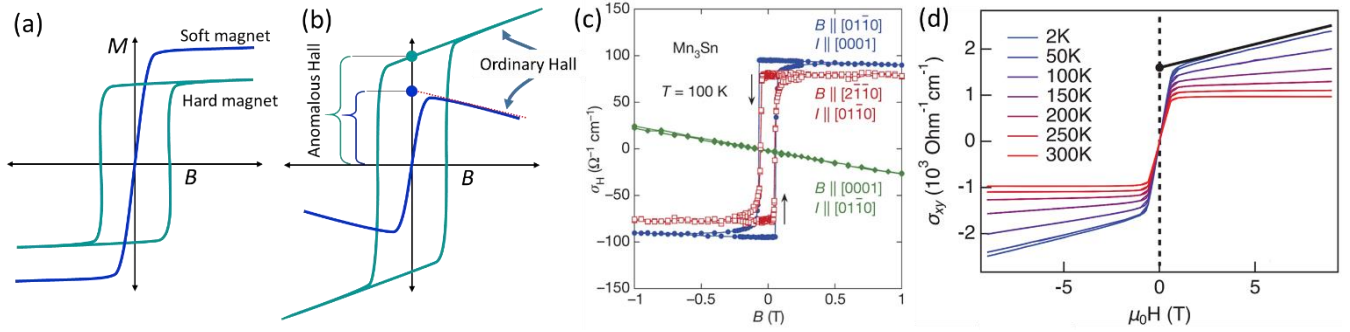

**Figure 20.** (a) Schematic of the field-dependent magnetization curve, *M*(*B*), for soft and hard magnetic materials. (b) Hall response curve as a function of the magnetic field for soft and hard magnetic materials. Field-dependent Hall conductivity (σ$_H$) plot with magnetic field for; (c) topological antiferromagnet Mn$_3$Sn, Reprinted with permission from ref. [34] Copyright 2016 Springer Nature, and (d) topological Heusler metal Co$_2$MnGa. Reprinted with permission from ref. [122] Copyright 2019 AAAS.

Another approach to achieve the quantized Hall effect in the absence of a magnetic field is the QAHE approach. Yu *et al.* proposed a theory for its experimental realization.[234] They demonstrated that if a ferromagnetically ordered state is achieved by doping the transition metal elements in the thin layers of three-dimensional topological insulators of Bi$_2$Te$_3$, Bi$_2$Se$_3$ and Sb$_2$Te$_3$, then the quantized Hall conductance ($\frac{e^2}{h}$) would be observed. This also requires that the Fermi energy lies in the magnetic gap, which can easily be achieved by electrostatic gating. Chang *et al.* verified this proposal in chromium-doped (Bi,Sb)$_2$Te$_3$ at low temperature[235] which was supported by other researchers.[236,237] As these systems are based on doping, even extremely careful considerations of the doping concentrations lead to the defects at the surface. Therefore, the precise quantization of the Hall conductance has not been observed above 2 K[238] (significantly below the magnetic ordering temperature), which demonstrates a major limitation of the QAHE at high temperature. Recently, an excellent solution was proposed for the antiferromagnetic van der Waals compound MnBi$_2$Te$_4$, which can be regarded as the intercalation of the MnTe layer within the quintuple Bi$_2$Te$_3$ layers.[239-241] MnTe and Bi$_2$Te$_3$ can function as antiferromagnetic and topological insulators, respectively. Deng *at al.*[242] experimentally demonstrated that it was possible to achieve the quantization of the Hall conductance up to 6.5 K by saturating all the spins through the application of a magnetic field.

Instead of the complex 2DEG, a recent exciting development is the QHE in single crystals.[194,243] Prior to this development, it was generally understood that all QHEs require two-dimensional semiconducting systems, which can support electrostatic gating. Bertrand Halperin had proposed, the possibility of a 3D QHE[244] from as early as 1987, but it was not until last year that researchers from Singapore observed the QHE in single crystals of ZrTe$_5$.[194] Recently, even a 3D version of the fractional QHE was observed in single crystals of the closely related compound, HfTe$_5$.[243] Physicists, materials scientists, and chemists can still dream, that we find other single crystalline materials with appropriate electronic structures (small anisotropic Fermi surfaces with instabilities), which exhibit a QHE at room temperature in the absence of a magnetic field, or even the QAHE in magnetic single crystals.

## 4.5. Anomalous Hall effect

In 1881, two years after the discovery of the Hall effect, Edwin Hall realized that ferromagnetic metals surprisingly exhibit a considerably larger Hall effect than nonmagnetic metals.[245] This observation was applicable for all ferromagnets and the phenomenon of the observation of the anomalously large Hall effect in ferromagnets became known as the anomalous Hall effect (AHE).[246] During the measurement, as the magnetic field increases, the Hall resistivity rises steeply at very small magnetic fields and saturates; after which, the change in the Hall voltage is very small (see Figure 20b). Notably, the overall Hall resistivity vs magnetic field curve resembles that of a typical magnetization vs magnetic field curve of a ferromagnet. This indicates that the anomalous Hall resistivity is proportional to the spontaneous magnetization (see Figure 20a and b). The Hall resistivity in the saturation region can increase or decrease with the field, depending on whether the material is *n*- or *p*-type, which means that it is in the region of the normal Hall effect as observed in the nonmagnetic metals. In order to quantify the AHE, the high-field normal linear Hall effect is extended to the zero magnetic field and the intercept in the Hall resistivity-axis is then identified as the anomalous Hall resistivity. In other words, the Hall resistivity at zero magnetic field is the anomalous Hall resistivity. It is also possible to obtain the carrier concentration of a material using the slope of the normal Hall resistivity region after saturation. The hard ferromagnetic material magnetization vs magnetic field curve exhibits a large hysteresis with remnant magnetization (see Figure 20b). The shape of the corresponding Hall resistivity vs magnetic field curve is similar and the anomalous Hall resistivity is equivalent to the remnant magnetization. After seeing the correspondence between the magnetization and Hall resistivity, it is obvious to write the Hall resistivity as $\rho_H = R_0 B + R_S M$. Here, the first term represents the normal Hall effect with $R_0$ being the normal Hall coefficient. The second term signifies AHE with $R_S$ being the material specific anomalous Hall coefficient. This means that, apart from the total magnetization, the extent of the AHE would largely depend on $R_S$. Importantly, the anomalous part is a product of the magnetization and $R_S$; hence a material with even a small magnetization but large $R_S$ or the vice versa can result in a large anomalous Hall effect.

The origin of $R_S$ has prompted considerable debate.[246] For many decades it was believed that $R_S$ and, hence, the AHE depended on the scattering of electrons from crystal imperfections that are completely extrinsic in nature. However, such mechanisms failed to explain the AHE in many good ferromagnetic metals where the conductivity remains in the intermediate range, i.e., it is neither a poor nor does it have a very large conductivity. In 1954, Karplus and Luttinger first proposed that AHE can be completely intrinsic in nature, without invoking imperfection-induced scatterings.[247] Based on this proposal, the electrons in a ferromagnet acquire a net anomalous velocity perpendicular to the direction of the applied electric field computed over all the filled states in the band structure. This net anomalous velocity that is perpendicular to the electric field can contribute to the Hall signal, even in the absence of the magnetic field. The net effect in a nonmagnetic or an antiferromagnetic metal is zero, thus yielding zero AHE. However, in a metallic ferromagnet, an intrinsic AHE is observed, which depends on the details of the band structure. It was later realized that the anomalous velocity is related to the Berry curvature, which should then also be computed over the whole filled stated to estimate the AHE. We will not focus on the Berry curvature; however, as a rule of thumb, the crossing points in the band structure contribute the most towards the Berry curvature. Although crossings are observable in the band structure of nonmagnetic or antiferromagnetic metals, the net Berry curvature is always zero because the negative Berry curvature cancels out the positive Berry curvature.[248]

It is now straightforward to ascertain the fact that magnetic topological semimetals are ideal candidates for the observation of large AHE with abundant crossing points in the form of Weyl points or in terms of nodal lines. In fact, a plethora of topological systems have already been established as large AHE materials due to hotspots in the Berry curvature in the vicinity of the crossing points. Non-collinear spin systems $Mn_3Sn$[34] and $Mn_3Ge$,[35,249] exhibit very large anomalous Hall conductivity values of the order of hundreds of $\Omega^{-1}cm^{-1}$, despite having magnetic moments that are only a few milli $\mu_B$/f.u. This deviation is due to the presence of multiple Weyl points (see Figure 20c).[36,37,162,250] This contrasts with some very strong ferromagnets with magnetic moments of up to 6 $\mu_B$/f.u, exhibiting weak AHE due to the small net Berry curvature.[248] This leads us to the very important conclusion that *it is possible to have a large AHE in a weak ferromagnet but a very small AHE in a strong ferromagnet,* all depending on the details of the distribution of Berry curvature from the occupied electrons in the band structure.[32,248] The large AHE in ferromagnetic topological systems $Co_3Sn_2S_2$,[30,31,251,252] $Co_2MnGa$,[32,122,253,254] and $Fe_3GeTe_2$[255,256] has been attributed to the presence of nodal line crossings close to the Fermi energy. These nodal line crossings are the source of the large Berry curvature (see Figure 20d for AHE in $Co_2MnGa$).

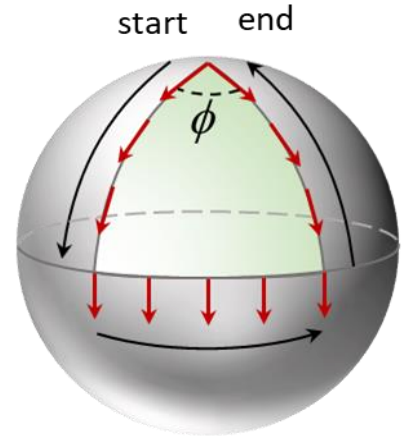

**Figure 21.** Schematic of a geometrical phase by parallel transport of a vector.

The Berry curvature distribution and, hence, the AHC can be tuned by modifying the crystalline symmetry or adjusting the chemical potential via suitable doping. An important constraint for observing the influence of Berry curvature or topological states on the physical properties of the associated material is that Fermi energy should be least populated with the non-topological bands. The parameter that can be assigned as a benchmark for the comparison is the anomalous Hall angle, which is estimated by $\Theta_{AHE} = \rho_{xy}^A/\rho_{xx}$, where $\rho_{xy}^A$ and $\rho_{xx}$ are the anomalous Hall resistivity and the longitudinal resistivity, respectively. Since the topological states dominate near the Fermi energy, the $\Theta_{AHE}$ value attains large values for $Co_3Sn_2S_2$ (20 %),[30] $Co_2MnAl$ (21 %),[171] and $Co_2MnGa$ (12 %).[32]

## 5. Berry phase and curvature

Berry phase is a geometrical phase which was first introduced by Sir Michael Berry for quantum states in 1984.[257] In Michael Berry's words the geometrical phase signifies a "global change without local change".[258] This can be best understood in terms of parallel transport of a vector on the surface of a sphere (curved surface) along a closed loop. Suppose (tangential to the surface of the sphere) a vector starts near the north pole, always pointing south and traverses a loop as shown in Figure 21.[259] When it reaches the starting point again, the vector points to a direction which makes an angle $\phi$ with respect to the starting vector. This additional acquired geometrical phase depends on the solid angle subtended by the closed loop. The Foucault pendulum is a physical example of a geometrical phase.[260] Michael Berry first showed that this concept also applies to a quantum mechanical state when it makes a loop in parameter space when traversed adiabatically and acquires a phase that is now referred to as Berry phase.[257] Zak realized that the Brillouin zone of a crystal makes a perfect parameter space of crystal momentum $k$, wherein the motion of an electron can be associated with the Berry phase.[261] The Berry phase can be calculated as the line integral of a potential termed the Berry potential or Berry connection in a closed loop.[204] Alternatively, the Berry phase can also be calculated as the surface integral of the Berry field or curvature in momentum space. Berry curvature can be viewed as an effective magnetic field in momentum space. Now that we understand that the Berry phase is defined in a closed loop, the most natural way one can envisage to make electrons move in a closed loop in the Brillouin zone is via a magnetic field. Mikitik and Sharlai demonstrated that the phase in the Shubnikov-de Hass and van Alphen-de Haas oscillation in metals in the presence of magnetic field contains the information of the Berry phase.[262] The Berry phase associated with a parabolic band is zero while it is $\pi$ for a linearly dispersing band. This was clearly demonstrated in graphene wherein the Dirac fermions occupying the linear bands acquire a $\pi$-Berry phase.[263,264] Recently, the Berry phase analysis of quantum oscillation measurements has become one of the major tools to differentiate topological Dirac or Weyl semimetals with linear bands from trivial semimetals with parabolic bands.[265,266] While the Berry phase depends on the closed path, the Berry curvature is a purely local quantity and therefore is more fundamental than the Berry phase itself.[204] Here, we mention two of the most common uses of the Berry curvature. First, it can be used to compute the topological charge (Chern number) of a point in the Brillouin zone for which one has to take the integral of the Berry curvature over a closed surface that encloses that point. For example for TaAs, the Chern number is ±1 depending on the chirality of the Weyl point.[99] Integrating the Berry curvature over a closed surface within the Brillouin zone that does not include a Weyl point gives zero. Thus, the Weyl points can be considered as a source or sink of the Berry curvature.[93] Second, the Berry curvature, can be used to compute the intrinsic anomalous Hall conductivity in a ferromagnetic material.[246] The anomalous velocity of electrons perpendicular to the applied electric field in zero magnetic field in a ferromagnet that was discovered by Karplus and Lutinger[247], was later realized to be related to the Berry curvature. Hence, the anomalous Hall conductivity is directly related to the net Berry curvature in a material.

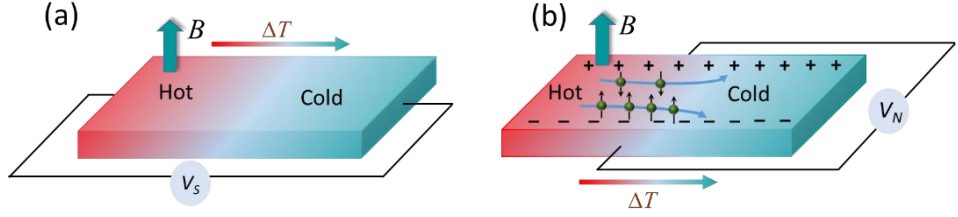

**Figure 22.** (a) Schematic representation of the Seebeck thermopower measurement. The longitudinal voltage ($V_S$) is produced by a temperature gradient ($\Delta T$). In case of magneto-Seebeck measurement the magnetic field is applied perpendicular to $\Delta T$. (b) Schematic representation of the Nernst measurement. In Nernst effect, a transverse voltage ($V_N$) is produced by a temperature gradient ($\Delta T$) and a magnetic field ($B$) orthogonal to each other. The application of a magnetic field results in the deflection of electrons and holes in the opposite transverse directions and contributes to the enhancement of the Nernst voltage of a dual charge carrier system.

## 6. Thermal transport in topological materials
### 6.1. Thermoelectricity

The electrons in solids not only carry charge but also heat, which forms the basis of thermoelectricity. Thermoelectric devices are all-solid-state devices that enable the direct conversion of heat to electricity or vice versa.[29,267-270] The beginning of the study can be traced back to as early as 1821 when German scientist Thomas Johann Seebeck discovered an interesting experimental result that the formation of a circuit with two non-identical bismuth and copper wires, each with a different temperature, would deflect the needle of a magnetic compass.[271] Current-carrying wires are known to generate a magnetic field. However, this was the first demonstration that a wire with a temperature gradient generates electricity by an effect known as the Seebeck effect. All the thermoelectric effects are related to heat that is transported by the charge carriers. Traditionally, the Seebeck effect has attracted considerable attention in an effort to improve the thermoelectric performance. The effectiveness of a thermoelectric material is governed by the dimensionless thermoelectric figure of merit, $zT = \frac{\sigma S^2 T}{\kappa}$, where $\sigma$ is the electrical conductivity, $S$ is the Seebeck thermopower, $T$ is the temperature in kelvin, and $\kappa$ is the thermal conductivity.[29,267-270] The fundamental challenge for designing a promising thermoelectric material is intriguing due to the conflicting thermoelectric parameter requirements. Since electrons conduct both electricity and heat, any attempt to increase the electrical conductivity of the materials often increases the thermal conductivity as well. To improve the thermoelectric properties, different strategies have been developed by improving the Seebeck thermopower, reducing the thermal conductivity, or simultaneously tailoring both parameters.[268,272-281] In general, the effective decoupling of electrical and thermal transport is crucial for the development of high-performance thermoelectric materials. The fundamental demands of similar material features, such as heavy elements and narrow band gaps connect the fields of topological insulators and thermoelectrics. In fact, the first-generation three-dimensional topological insulators were discovered in very well-known thermoelectric systems $Bi_{1-x}Sb_x$[63] and $(Bi,Sb)_2(Te,Se)_3$.[67] For a more detailed discription of the commonalities of topological materials and thermoelectrics, see refs[28,282-285].

### 6.2. Seebeck effect

Seebeck coefficient or Seebeck thermopower is equivalent to the resistivity. However, instead of applying a current, a temperature difference between two ends of the sample is applied and the voltage generated is measured in the same direction. Figure 22a depicts a typical representation of the Seebeck measurement. The Seebeck thermopower is the ratio of a resulting electrical voltage to an applied temperature gradient. The quantity is measured by stabilizing an exact temperature difference (steady-state) or by continuously measuring while the temperature difference is varied slowly (quasi-steady-state).[286] The sign of Seebeck thermopower denotes the dominant carrier in a particular material. A positive sign of the *S* indicates *p*-type conduction i.e. hole is the dominant charge carrier in the system. On the other hand, a negative sign denotes *n*-type conduction i.e. electron is the dominant carrier. Although the exact temperature dependence of the Seebeck thermopower is highly material-dependent, Figure 23a depicts representative response curves with respect to temperature. Curves i and ii are negative and positive, respectively, in the entire temperature range, thus signifying electron and hole dominated transport, respectively. Alternatively, a single material can exhibit switching of the carrier type with temperature, as represented by curve iii, among many other possibilities (e.g., *p-n-p*, *n-p*, and *n-p-n*).[292-294] The Seebeck thermopower is usually small for metallic materials, since both electron- and hole-like charge carriers will be thermally excited and have opposite contributions to the Seebeck thermopower, thus compensating each other. Alternatively, a semiconductor generally exhibits a large Seebeck thermopower. In most cases, a heavily doped semiconductor with a carrier concentration in the range of $10^{19}$–$10^{21}$ cm$^{-3}$ is a good selection for thermoelectric applications,[29,267] where a Seebeck coefficient in the range of 100–300 $\mu$V/K could be achieved.

Table 2. Anomalous transport properties of magnetic topological materials

| Compound | Topology type | AHE (~2K) ($\Omega$ cm) | AHC (~2K) ($\Omega^{-1}$ cm$^{-1}$) | AHA (maximum) (%) | ANE ($\mu$V K$^{-1}$) | Ref. |
|---|---|---|---|---|---|---|
| GdPtBi | Weyl | $1.8 \times 10^{-4}$ | 200 | 16 (2K) | - | 287 |
| Co$_3$Sn$_2$S$_2$ | Weyl | $2.8 \times 10^{-6}$ | 1130 | 19.8 (75K) | 3 (80 K) | 30,40 |
| Co$_2$MnGa | Nodal line | $5.0 \times 10^{-6}$ | 1600 | 12 (300K) | 6 (300 K) | 32,39 |
| Co$_2$MnAl | Nodal line | $3.7 \times 10^{-5}$ | 2000 | 22 (2K) | - | 171 |
| Fe$_3$Sn$_2$ | Dirac with a gap | - | 1050 | - | - | 288 |
| Fe$_3$GeTe$_2$ | Nodal line | $1.3 \times 10^{-5}$ | 540 (10K) | 9 (10K) | 0.3 (50 K) | 255,289 |
| Fe$_3$Ga | Nodal line | $3 \times 10^{-6}$ | 610 | - | 6 (300 K) | 290 |
| Fe$_3$Al | Nodal line | $4 \times 10^{-6}$ | 460 | - | 4 (300 K) | 290 |
| Mn$_3$Sn | Weyl | $4.15 \times 10^{-6}$ (100 K) | 100 (100 K) | 3.2 | 0.6 (200 K) | 34,38 |
| Mn$_3$Ge | Weyl | $5.1 \times 10^{-6}$ (100 K) | 500 | 5 | 1.5 (100 K) | 35,291 |

In a topological insulator system, the overall sign and nature of the Seebeck thermopower is governed by the relative contribution from the surface and bulk band. The strong energy dependence in the scattering time caused by the bulk-boundary interactions may lead to an unusual transport property like the opposite dominant carrier type in the thermal and electrical transport property, which is known as the anomalous Seebeck thermopower.[285,295] Thus, the measurements of electrical and Seebeck response could help to elucidate the bulk and topological surface states of topological insulators. For instance, quintuple-layer (Bi$_{1-x}$Sb$_x$)$_2$Te$_3$ films exhibit a sign anomaly between the Hall coefficient (R$_H$) and *S*.[296] The effect of topological surface states on the thermoelectric transport is a rather unexplored area of research. Therefore, the exploration of the potential of topological surface state dominated thermoelectric transport is an exciting field, wherein, topological surface states can simultaneously exhibit sizable *S* and superior mobility, which are both advantageous for thermoelectric applications. In this context, the investigation of nanostructured topological insulator materials will also be interesting for exploring the effect of topological surface states. Promising thermoelectric properties have been achieved in topological insulators Bi$_{1-x}$Sb$_x$,[297] layered tetradymites Bi$_2$Te$_3$,[298-304] Sb$_2$Te$_3$,[305-307] alloys of Bi$_2$Te$_3$-Sb$_2$Te$_3$,[308-310] Bi$_2$Se$_3$[311-313], topological crystalline insulator SnTe,[314-317] and Pb$_{1-x}$Sn$_x$Te.[318] In general, the topological surface state-dominated thermoelectric transport in bulk and nanostructured topological insulators provide a new frontier in thermoelectric research.

Most semimetals including topological Weyl and Dirac semimetals carry both electrons and holes as charge carriers. In the presence of a longitudinal temperature gradient, both the holes and electrons move to the same side of the sample, thereby resulting in very small Seebeck voltages as they counterbalance each other's contribution towards the net Seebeck thermopower. Generally, the transport properties of topological semimetals are dominated by linear band dispersions. As discussed previously, linear band crossing results in very few charge carriers, small effective mass, and very large carrier mobility and magnetoresistance. Therefore, the magneto-thermoelectric measurement can be used to access the topological band structure effects and the enhanced Seebeck thermopower. A recent theoretical study indicates a large non-saturating thermopower in Dirac and Weyl semimetals in the magnetic field.[319] Magneto-thermoelectric measurements have been conducted in Dirac semimetal Pb$_{1-x}$Sn$_x$Se,[320] PtSn$_4$,[190] and Cd$_3$As$_2$[321-323]. The Dirac semimetal Cd$_3$As$_2$ exhibits improved zT values from 0.23 to 1.24 at 450 K in a magnetic field of 9 T due to the enhanced Seebeck thermopower.[322] However, until now, the experimental and theoretical studies in this direction are limited. Thus, further research is required to understand the potential of topological semimetals for magneto thermoelectrics based on the Seebeck effect. Another potential area is transverse thermoelectrics where, the topological semimetal could be an excellent candidate, which we discuss in the next section.

## 6.3. Nernst effect

The configurations based on the Nernst effect are known as the transverse thermomagnetic effect. Here, a transverse electrical signal is generated in the presence of a mutually perpendicular magnetic field and temperature gradient.[324,325] This can be viewed as a Hall voltage measurement with a temperature gradient instead of an electric current. Therefore, the magnetic field-dependence response curve of the Nernst thermopower is comparable to the field dependence of the Hall resistivity curve. Historically, the Nernst effect attained significant research attention after the discovery of a large Nernst signal in a high-temperature cuprate superconductor.[325,326] Thereafter, many heavy-fermion systems and superconductors have been studied. More recently, increasing attention has been placed on the study of the Nernst effect in topological materials. In a magnetic field orthogonal to the temperature gradient, the electrons and holes become deflected to the opposite transverse direction due to the Lorentz force. Therefore, it is possible to achieve a large Nernst signal in a dual charge carrier or compensated topological system. Compared to the traditional thermoelectrics based on the Seebeck effect, those based on the Nernst effect have generated much less attention. The multi-terminal thermoelectric devices based on the Nernst effect enable the spatial separation of the heat reservoir from the electric circuitry. Moreover, both *p*- and *n*-type materials are not required in Nernst devices because the magnetic field direction can reverse the polarity of the voltage.[183,327,328] As one-type (*p*- or *n*-type) materials are sufficient for the devices, the Nernst devices overcome the compatibility issues of *p*- and *n*-type materials that arise from the different thermal expansion coefficients. In this section, we first discuss the ordinary Nernst effect and then the anomalous Nernst effect.

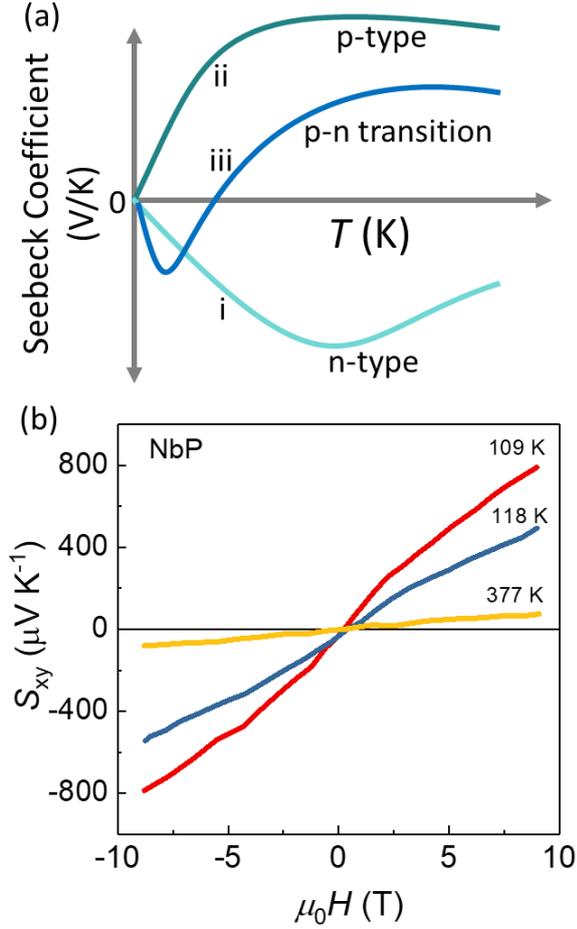

**Figure 23.** (a) Schematic representation of the different types of Seebeck response curves as a function of temperature. (b) Magnetic field dependence of Nernst thermopower ($S_{xy}$) for NbP single crystals at different measurement temperatures. Adapted with permission from ref. [183] Copyright 2018 American Physical Society.

The signal response curve for an ordinary Nernst effect increases linearly as the strength of the magnetic field increases. This behavior is generally observed in non-magnetic systems with one type of dominant charge carrier. Figure 23b depicts the field dependence of the Nernst thermopower for single crystalline NbP at different temperatures. The Nernst thermopower increases almost linearly and exhibits an unsaturated trend with the increase in the magnetic field. The single crystalline Weyl semimetals NbP exhibited a very large Nernst thermopower of approximately 800 µV K$^{-1}$ at 109 K and 9 T, which is two orders of magnitude higher than the Seebeck thermopower.[183] The combined experimental and theoretical study shows that as temperature increases, the chemical potential shifts to the Dirac-point, which enhanced the Nernst effect. A large Nernst effect was also observed in polycrystalline NbP.[327] The studies have also been extended to other non-magnetic topological candidates, such as type-II Weyl semimetal WTe$_2$[186] Dirac semimetal Cd$_3$As$_2$,[188,329] Pb$_{1-x}$Sn$_x$Se,[320] and ZrTe$_5$.[330,331] All these non-magnetic topological semimetals exhibit promising Nernst thermopowers. A Nernst thermoelectric figure of merit $zT_N$ (~0.5) has been achieved in Cd$_3$As$_2$ at room temperature, due to moderately high Nernst thermopower and intrinsic low thermal conductivity.[188] Although the latest investigation is promising, the studies on the Nernst effect is limited for non-magnetic topological semimetals. Therefore, further exploration is required to effectively contribute to thermoelectric research. There are a few key points that should be considered for the selection of materials: (a) topological semimetals with high charge carrier mobility and low carrier effective mass, (b) topological semimetal materials with low phonon group velocity so that intrinsically low thermal conductivity can be achieved to maximize figure of merit, (c) the large signal should be observed in low magnetic fields so that the device can be operable using a low magnetic field, and (d) the signal should be close to room temperature or above room temperature for power generation applications.

### 6.4. Anomalous Nernst effect

A high external magnetic field is generally required for observing large ordinary Nernst signals. Anomalous transverse voltages in ferromagnetic materials generate mutually perpendicularly to both temperature gradient and magnetization, which is termed the anomalous Nernst effect (ANE). In contrast to ordinary Nernst signals, anomalous Nernst signals rapidly increase at low magnetic fields, as demonstrated in Figure 24a. Therefore, large Nernst signals are achievable at lower magnetic fields, which is a major advantage. Moreover, ANEs can even be observed in hard magnetic systems at zero magnetic field, which is extremely important for Nernst thermoelectric devices as no external magnet is required (see Figure 24b).[40] Anomalous Nernst signals in conventional magnetic materials scale up linearly to the magnetization of materials. However, this relationship is not valid in topological systems with a Berry curvature effect and the anomalous signal exhibits beyond the magnetization scaling.[332] Figure 24c depicts the Nernst signal of conventional ferromagnets and topological magnets. All the topological magnets such as, Co$_2$MnGa,[333,334] Co$_3$Sn$_2$S$_2$,[39,40,252,335] Fe$_3$Al,[252] and Fe$_3$Ga[252] exhibit significantly higher signals than conventional materials. This effect has also been observed in the antiferromagnetic topological compound Mn$_3$Sn,[38,336] and Mn$_3$Ge.[291] Notably, the anomalous signal observed for topological magnets is considerably smaller than the threshold requirement of thermoelectric applications. Therefore, the search for new topological magnets is of extreme importance, particularly those with high Curie temperatures. The additional characteristics should be considered for the material selection: (a) topological magnets with high Curie temperature, which enables the high ANE signal to be achievable close to room temperature or high temperature. (b) Low magnetic moment will be particularly useful for eliminating the stray field in the device. (c) Hard magnets are more favorable than soft magnets for obtaining the zero-field anomalous Nernst thermopower.

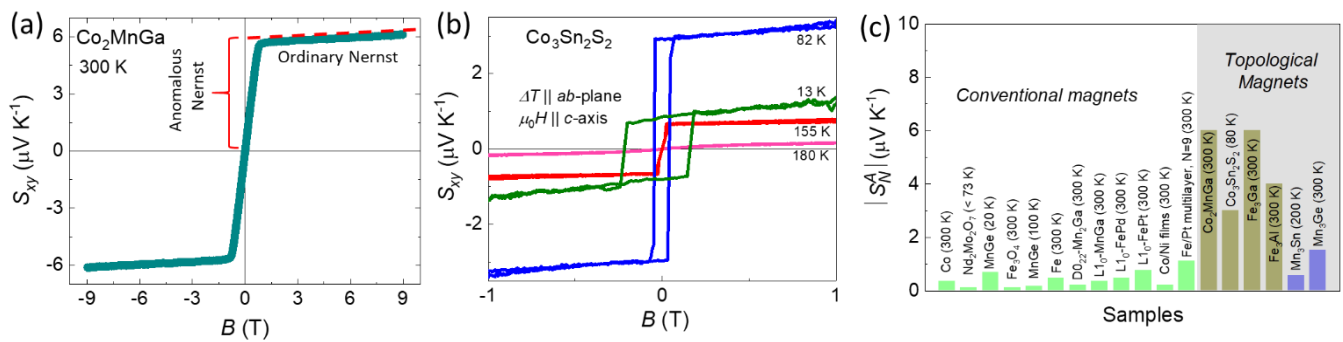

**Figure 24.** (a) Magnetic field-dependence of the Nernst thermopower ($S_{xy}$) of $Co_2MnGa$ at 300 K. The ordinary and anomalous response contributions to the Nernst thermopower can then be separated by fitting the high-field slope of the total Nernst thermopower, as indicated by the middle bracket and dotted line in the plot. Reprinted from ref. [39] with permission under CC BY 4.0 license. Copyright Springer Nature. (b) Magnetic field-dependence of Nernst thermopower for hard magnet Weyl semimetal $Co_3Sn_2S_2$ at different temperatures. Reprinted from ref. [40] with permission under CC BY-NC 4.0 license. Copyright WILEY-VCH Verlag GmbH & Co. KGaA, Weinheim. (c) Absolute value of anomalous Nernst thermopowers for conventional and topological magnets.

## 7. Topology in oxides

An interplay between the charge, spin and orbital degrees of freedom imparts exciting electronic and magnetic properties in oxides. Transition metal oxides, in particular, have been at the forefront of many exotic phenomena in condensed matter such as high temperature superconductivity[337-341], multiferroics[342,343] and colossal magnetoresistance[344]. Since the advent of topology in solid state materials, there have been efforts to study the topological degree of freedom in oxides. Initial research on topological insulators outlined the importance of large spin orbit coupling for generating large enough inverted band gaps for room temperature applications. It was a natural question to ask then, whether the inclusion of electronic correlation, which is abundant in oxides, can bring novel topological features.[345] Perovskite $BaBiO_3$ was the first predicted non-transition metal oxide topological insulator on appropriate electron doping.[346] Although, the inverted band gap (0.7 eV) which hosts the surface Dirac cone is quite large, the main challenge lies in achieving the sufficient electron doping to shift the gap to the Fermi energy. Among transition metals, correlation is maximum in 3$d$-transition metals while it decreases in 4$d$- to 5$d$- transition metals because the $d$-orbitals become increasingly extended. On the other hand, spin orbit coupling increases from 3$d$- to 5$d$-transition metals. In 5$d$-transition metal oxides the strength of electronic correlation and spin orbit coupling is of the same order which make them excellent candidates to realize novel topological insulators. Shitade et al. predicted a two dimensional topological insulating state in the paramagnetic phase of Mott insulator $Na_2IrO_3$.[347] Later, ARPES studies carried out by Alidoust et al. demonstrated the in-gap metallic surface state confirming the topological insulating state in this compound.[348] Wan et al. showed that the pyrochlore iridate $Y_2Ir_2O_7$ might contain Weyl semimetallic states at large correlation and all-in/all-out antiferromagnetic spin structure.[93] In this spin structure, the corner-shared tetrahedra of iridium atoms contain spins pointing all inward and all outward maintaining the centre of inversion in the structure. A clear evidence of a topological state in this compound is still lacking, partly due to the unavailability of high quality single crystals. Interestingly, Juyal et al. observe a charge density wave instability and chiral anomaly induced negative magnetoresistance in single crystalline nanowires of $Y_2Ir_2O_7$ which hints towards the gapping of the Weyl nodes.[349] The first three dimensional Dirac semimetal was predicted in $BiO_2$ assuming that the structure is same as that of β-crystobalite $SiO_2$.[350] Unfortunately, this assumed structure is thermodynamically unstable, and therefore could not be verified in experiments. Rutile type $IrO_2$, $OsO_2$ and $RuO_2$ are known to be highly metallic.[351,352] Sun et al. showed that these compounds in the absence of spin orbit coupling contain long Dirac nodal lines.[353]

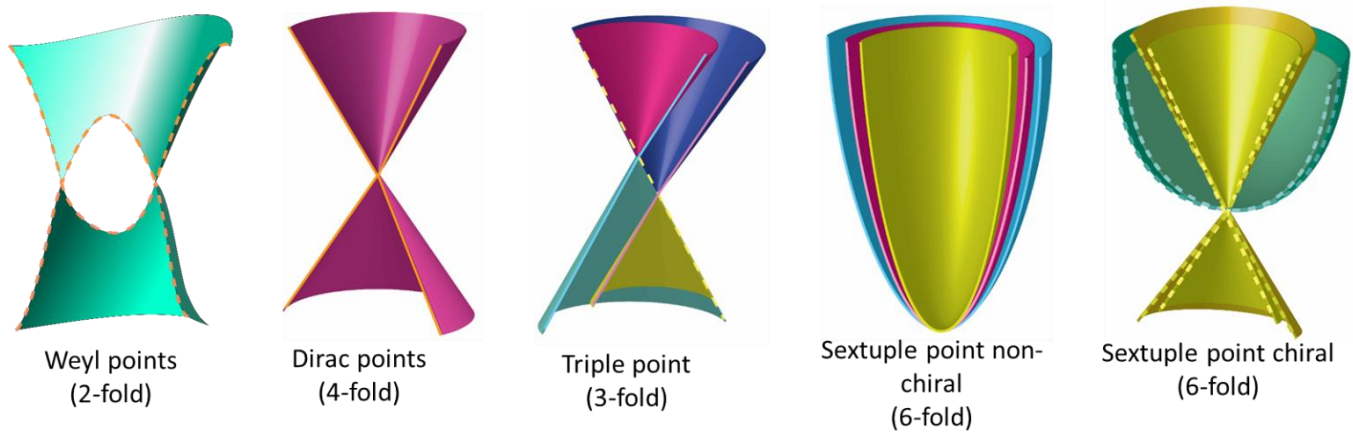

**Figure 25.** Schematic of the differently degenerate topological crossings in 2-fold Weyl, 4-fold Dirac, 3-fold triple point, 6-fold sextuple point in achiral pyrites and 6-fold sextuple point in chiral silicides.

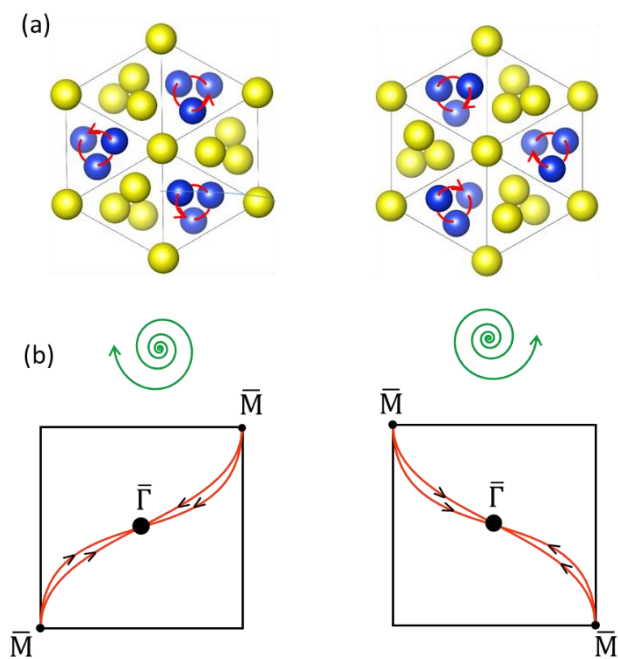

**Figure 26.** (a) Crystal structure of the chiral binary compounds in the B20 structure of opposite chirality and (b) their corresponding Fermi arcs.

The bands at the nodal lines stick together due to the presence of mirror and glide planes in the crystals. On the application of spin orbit coupling a band gap is generated which is the origin of the strong spin Hall effect in these compounds.[353,354] In order to search for large spin Hall effect materials the authors prescribe to look for systems with high symmetry containing a number of mirror and glide planes.[353]

## 8. New Fermions

Dirac and Weyl particles were originally predicted to occur as free particles in space rather than on the inside as solid crystals. Electrons inside solid matter can only mimic the properties of the proposed Dirac and Weyl particles due to various symmetries, i.e., crystalline and time reversal symmetry/presence or absence of magnetism. Then, are only Dirac and Weyl-like particles in crystalline solids? Bernevig and coworkers demonstrated that it is possible to obtain a variety of other quasi-particles in solids that otherwise cease to exist as free particles in space due to less severe symmetry constrains in the solid state matter compared to the high energy physics[355]. For example, there can be three-fold, six-fold, or eight-fold degenerate band crossing points protected by certain crystalline symmetry elements. This classification has been conducted for nonmagnetic compounds. The three-fold degenerate point or triple point results from the crossing of a doubly degenerate band with a non-degenerate band, as shown in Figure 25. The initial proposal for the triple point was limited to crystals with nonsymmorphic symmetry where the rotation or mirror operation is combined with a lattice translation of a fraction of the primitive unit cell vector. Later, Soluyanov and co-workers highlighted several examples of symmorphic crystal systems where triple points can be stabilized.[356] Figure 25 depicts that if one of the doubly degenerate bands is split in a Dirac crossing, it results in a pair of triple-point crossings. Alternatively, if the doubly degenerate band of triple point crossing is split by the application of the magnetic field, then a pair of triple points gives rise to two pairs of Weyl points. Ding and coworkers conducted ARPES measurements on hexagonal systems (MoP[357] and WC[358]) and provided experimental evidence of the triple point. Interestingly, MoP[149] exhibits a more superior low-temperature conductivity than copper metal and WC[359], in the presence of a magnetic field, suggests a transport that is influenced by Weyl points. ZrTe[360] is another candidate triple point compound that exhibits a unique power dependence of $MR$ along the crystallographic axis where the triple points are proposed to reside. The sixfold degenerate point in the band structure can have various types. In compounds with pyrite structures like superconducting $PdSb_2$, three two-fold degenerate bands touch at the corner of the

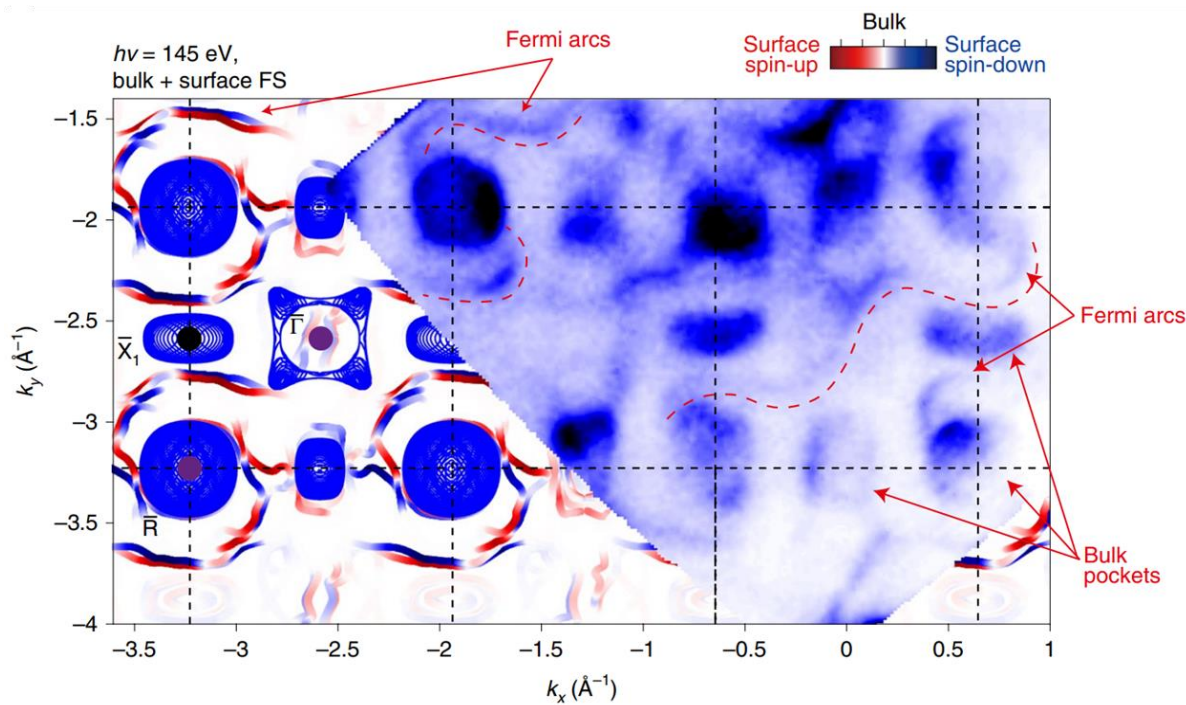

**Figure 27.** ARPES data and corresponding ab-initio calculations of structurally chiral topological compound AlPt showing long Fermi arc extending the whole Brillouin zone. Reprinted with permission from ref. [369] Copyright 2019 Springer Nature.

Brillouin zone to yield a six-fold degenerate point or sextuple point. Kumar *et al*. provided signatures of this degeneracy with the help of ARPES.[361-363] The sister compound PtBi$_2$ also contains the sextuple point, but in the unoccupied conduction band. The extremely large MR in this compound has been attributed to the presence of independent Dirac points in the occupied valence band.[192]

Another new type of gapless electronic excitations forms in chiral topological compounds with several chiral crystal structures.[364,365] Here, the atoms arrange in a spiral staircase pattern, as shown in Figure 26a. The special crystalline symmetry of the chiral structure protects these symmetry-enforced multifold band crossings at the high symmetric points of the Brillouin zone. Two multifold crossings are observed in the chiral crystals CoSi,[140,141,366,367] RhSi,[141,368] and AlPt[369] at the zone center Γ point and the corner point R of the Brillouin zone. A new type of four-fold degenerate chiral crossing and a six-fold degenerate point forms at Γ and R, respectively. Both of these multifold points carry a topological charge (Chern number) of ±4, and are thus variants of the Weyl point connected by the Fermi arc expanding over the entire Brillouin zone (see Figure 26b). This Fermi arc is longer than any other compounds investigated thus far (see Figure 27 for experimental data of AlPt). The quantized topological charge, or the Chern number of these Weyl fermions (±4) is four times larger than any conventional Weyl fermionic system and is the maximal value possible in nature.[365] This is a distinctly different scenario than other multifold crossings discussed earlier, where the topological charge is zero.

The topological charge of a Weyl point is determined by counting the net crossing of the surface states across an enclosed loop around the Weyl point. The maximal quantized charge, i.e., 4 of the chiral multifold fermions, has recently been verified in compounds[370] and PdGa[165]. These compounds exhibit higher SOC strengths than CoSi or RhSi-like compounds, which is helpful for increasing the Fermi arc splitting and easily determining the Chern number.[165,370] Since the multifold fermions are strongly bound and protected by the crystalline symmetry, a change in the handedness of the chiral structure, i.e., left-handed chiral to right-handed chiral, also changes the associated Fermi arc velocities.[165,166]

The multifold crossing with maximal quantized charge in chiral semimetals results in many exotic phenomena like chiral the magnetic effect, quantized circular photogalvanic effect and other optoelectronic phenomena. The mirror symmetries in the chiral $P2_13$ (No. 198) space group are replaced with the glide operation. Consequently, the two Weyl points reside at different energy values in the chiral crystals. This is a different scenario than that of conventional Weyl semimetals like NbP or TaAs, where the Weyl points form at the same energy due to the mirror symmetry of the crystal structure.[99] As a result, in the optical excitation, one Weyl point can be Pauli blocked, whereas the other divulges quantized *dc* current in the circular photogalvanic experiments.[161,365,371]

## 9. Nonlinear optical responses

Nonlinear optical response in materials is one of the most active research areas today because of technological applications in the area of energy applications such as the bulk photovoltaic effect in solar cells. The term nonlinear means that on application of intense light on a material, the induced electrical polarization or the current is a nonlinear function of the electrical field. Some of the most important nonlinear optical responses are second harmonic generation (SHG),[372] the circular photogalvanic effect (CPGE) [373-375] and the bulk photovoltaic effect[376-378]. Since the even order optical phenomena are forbidden in systems with inversion centres, noncentrosymmetric materials are required to observe the above mentioned second order nonlinear optical effects. Hence, nonmagnetic Weyl semimetals which must break the centre of inversion have recently caught a lot of attention in this field. Wu et al. showed that the Weyl semimetal TaAs is one order of magnitude more effective in generating second harmonics than the previously known benchmark material GaAs.[379] In SHG, two photons of the same frequency $\omega$ interact with the active medium to give rise to a single photon of double the frequency $2\omega$. An important observation in this respect is that the crystal structure of TaAs allows one to define a polar vector which in the case of insulators like BaTiO$_3$ can result in spontaneous polarization (ferroelectricity). In contrast, due to the rotation symmetry in the noncentrosymmetric GaAs, such a polar axis does not exist. Although the topological nature of TaAs which allows for the finite net Berry connection between the two bands involved in the optical excitation process can be invoked to understand large SHG, a clear connection has not yet been established. However, this finding is certainly a motivation for studying SHG in nonmagnetic Weyl semimetals. CPGE is a second order nonlinear optical response, whereby a circularly polarized light can generate a photocurrent.[373-375] Chan *et al*. predicted that in Weyl semimetals the optical excitations in the bands corresponding to the Weyl points of opposite chiralities produce photocurrents propagating in opposite directions.[380] Since the Weyl points comes in pairs of opposite chiralities, the net photocurrent must be zero. In order to make an imbalance between photocurrents of Weyl points of opposite chiralities, the Weyl cones should have a finite tilt with respect to the energy-axis. Ma et al. measured the photocurrent in the Weyl semimetal TaAs by shining circularly polarized mid-infrared light.[10] They attribute this observation to a finite tilt of the Weyl cones with respect to the energy axis. Going beyond just measuring the photocurrent in Weyl systems, Moore and coworkers predicted that for certain Weyl semimetals CPGE induced current can be quantized to a value which depends only on the topological charge of the Weyl point and the intensity of the circularly polarized light used.[371] However, in order to show the quantization, Weyl semimetals should have low crystal symmetry with no mirror planes such that the Weyl points of opposite chiralities lie at different energies. It argues that for a range of frequency, only one Weyl point will participate in the inter-band photoexcitation to generate photocurrent while such excitation will be Pauli blocked (excitation to an already occupied state) for the Weyl point of the opposite chirality. Rees *et al*. studied CPGE in RhSi, which is a Weyl semimetal with structural chirality, hence the Weyl points are separated in energy with a topological charge of ±4.[161] Consequently, a large CPGE amplitude is observed in the predicted range of the photon energy. Outside this range, the amplitude decreases rapidly because Weyl points of both the chiralities contribute and the net effect is zero. Although a perfectly quantized CPGE was not observed in RhSi mainly due to the non-Weyl band present at the Fermi energy. Hence, it is important to search for Weyl semimetals with simpler band structures in chiral crystal structures.

## 10. Topological surface states

The surface states of the topological materials are of prime importance because they are central to many fundamental phenomena and carry potential for various applications. These surface states in topological materials are as a result of the bulk electronic structure. Therefore, they do not depend on the dangling bond scenario, as in the case of a trivial semiconductor. In this respect, the topological surface states are robust, which is effectively expressed in the experiments. The topological surface state in a three-dimensional topological insulator is two-dimensional, thus covering the entire surface of the material. When observed with respect to the energy, the surface states form a Dirac

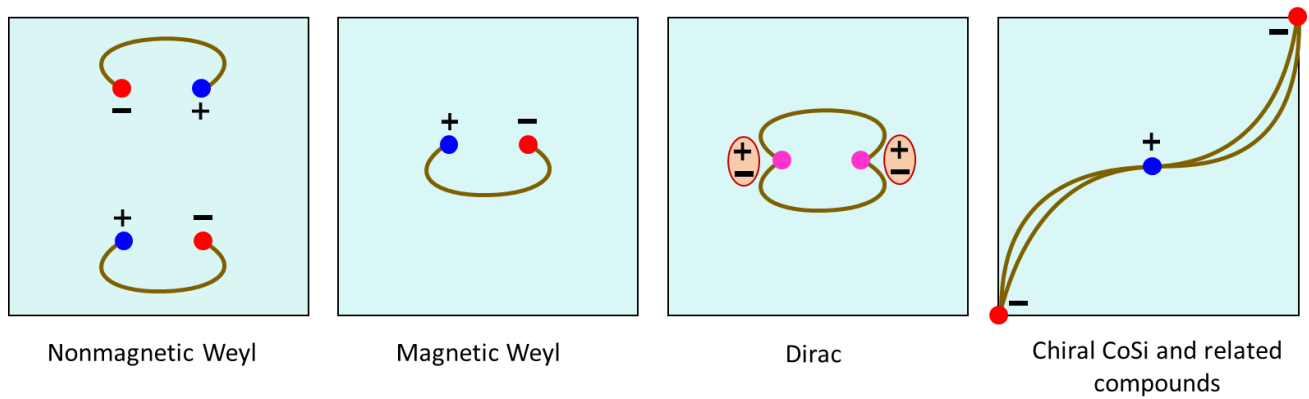

**Figure 28.** Schematic of the topological surface state Fermi arc in the nonmagnetic Weyl semimetal, magnetic Weyl semimetal, Dirac semimetal, and chiral CoSi and related materials

cone. The surface states in Weyl semimetals are special. Unlike any conventional surface states, the aforementioned occurs in unclosed arcs, known as Fermi arcs. The starting point of the Fermi arc is the surface projection of the Weyl point of one chirality in the bulk to the surface projection of the corresponding Weyl point of opposite chirality (see Figure 28). Hence, the length of the Fermi arc depends on the separation of the Weyl points of opposite chirality in the bulk. In the TaAs series of Weyl semimetals, the distance between the two Weyl points of opposite chirality increases as the SOC increases. Therefore, the length of the Fermi arc is largest in TaAs and smallest in NbP.[105] This cannot be generalized and might not hold true for other Weyl systems like transition metal dichalcogenides WTe$_2$ and MoTe$_2$ in their orthorhombic crystal structure (Td-phase). Pairs of the Weyl points appear in these two compounds due to sidewise touching of the conduction and valence bands. The distance between the two Weyl points and the length of the Fermi arc is considerably larger in MoTe$_2$ than in WTe$_2$, despite MoTe$_2$ being lighter than WTe$_2$.[109,113] The distance between the two Weyl points of opposite chirality in the momentum space signifies their stability because the closely situated Weyl points annihilate to lose their topological effect. The separation of the Weyl points in TaAs and the WTe$_2$ family of Weyl semimetals is only a small fraction of the Brillouin zone.[105,113] Yazyev and coworkers proposed robust Weyl points in noncentrosymmetric orthorhombic compounds WP$_2$ and MoP$_2$.[381] The neighboring Weyl points are of same chirality. However, the Weyl points of opposite chirality are well separated, thus providing protection against annihilation with each other. For the same reason, the Fermi arcs are long and extend a large portion of the Brillouin zone.[381]

An interesting question arises whether Fermi arcs also exist in Dirac semimetals. Since a Dirac point can be assumed to contain topological charges of +1 and -1 (net charge = 0), then it is possible to connect this point with another Dirac point containing -1 and +1, via two Fermi arcs, which when considered together gives rise to a closed loop. However, it has been argued that the Fermi arcs in Dirac semimetals are not topologically protected because of the lack of a net topological charge on a Dirac point.[382] Similarly, the Fermi arcs have also been observed in the triple point fermion metal WC. However, the lack of topological charge on these triple points cease to provide topological protection to the Fermi arcs.[358]

The Fermi arcs are expressed most effectively in the structurally chiral topological system CoSi and related compounds. Multifold degenerate points with topological charges of +4 and -4 exist at the center and the corner of the Brillouin zone. Hence, the Fermi arcs span the entire Brillouin zone. It is possible to obtain the topological charge of the point in the Brillouin zone by counting the total number of Fermi arcs originating or terminating from it. However, the ARPES experiments reveal that it is difficult to resolve separate Fermi arcs in systems with low SOC due to the limited resolution.[140,141,366,367] Yao *et al*. clearly demonstrated the splitting of the Fermi arcs by considering a system PtGa with large SOC, thus making the abovementioned counting rule observable.[370]

## 11. Outlook and future directions

The first question that comes up is whether there is still science to do in the field of topology, especially from the perspective of chemistry. The clear answer is yes: we have only seen the tip of the iceberg. It has been calculated that, of all the known inorganic compounds that can be described within a single particle picture, more than 20% are topological. So far only a tiny number of oxides have been investigated: oxides and other highly ionic compounds are not well described by singe particle methods based on the local density approximation. Several of the predicted topological oxides are highly insulating, so much so that their transport properties cannot be measured, and others that were predicted to be semiconductors or semimetals, had charge carrier concentrations more like a metal. In physics the disagreement between predictions and experiments are often related to electron-electron correlations. Some examples of predicted topological materials are rare earth and actinide pnictides[383], filled Skutterudites[384] and BaBiO$_3$.[346] More elaborate theoretical methods – and longer computing times – are necessary to investigate oxides, fluorides and nitrides. However, since these materials have larger band dispersions and larger band gaps, more stable and more interesting topological materials can be expected. It is also necessary to grow single crystals of Dirac, Weyl and new Fermions with very low defect densities. It took years before Bi$_2$Se$_3$ could be synthesized with a low charge carrier concentration in the bulk. In this class of materials their properties can be improved by defect compensation.[91] In systems with Weyl or Dirac points it is necessary that the Fermi energy is located exactly (in the meV regime) at the desired position in the band structure. For the TaAs family, several types of defects have already been identified.[385] A strategy can be to carry out more sophisticated electronic structure calculations, taking into account correlation effects, disorder, defects *etc*., and to synthesize more and better single crystals to unravel more intrinsic properties of topological materials. With the knowledge of vast materials classes, structural correlations among them, compositional tuning and defect engineering, solid state chemists in particular can enrich the field of topological materials to a great extent. For example, topology in oxides which is rather an unexplored area of research can bring the field to the next level with the emergent correlated topological phases.

Another obvious next step is the systematic investigation of magnetic materials. This is easy for ferromagnets: a crossing point in the band structure of a centrosymmetric ferromagnet is a Weyl point. Antiferromagnets are more complicated but can serve as model systems in astrophysics since they might host axion quasiparticles[240,386]. The general approach of "topological quantum chemistry" must be extended to magnetic space groups. In general, topological materials can serve as model systems for proposals in high energy and astrophysics and beyond, as we will learn from the new Fermions[355]. The difference of the new Fermions in the Universe and in a crystal was already recognized by Heisenberg: "In a Gitterwelt (lattice world) hosted electrons that could morph into protons, photons that were not massless, and more peculiarities that compelled him to abandon "this completely crazy idea"[387,388].

Additionally, several proposals have been made to use Weyl and Dirac semimetals in detectors for dark matter.[389] So far dark matter detector materials have only allowed the investigation of a small energy range, whereas topological materials will enable the investigation of a much larger energy range: topological materials could improve state-of-the-art detector sensitivity to energies smaller than 100meV. They also might lead to new infra-red detectors and cameras without the need of any cooling.

However, the most exciting direction for us as chemists are topological photovoltaics, topological thermoelectric applications and topological catalysis. Moore's team has developed a model to calculate shift currents for photovoltaics beyond the Shockley–Queisser limit of conventional solar cells.[390] This non-linear optical response can be strong in non-magnetic Weyl semimetals and chiral crystals, and, moreover, such devices will be simpler since the pertinent effect is a bulk photovoltaic effect.[391,392] Topological materials allow for going beyond the Wiedemann-Franz law that limits the thermoelectric figure of merit.[28] The thermal transport of topological materials can be also enhanced in a magnetic field, the so-called magneto-Seebeck effect, the Nernst effect, and the anomalous Nernst effect have the potential for efficient energy conversion.[11]

New topological quantum technologies are another significant future direction which might lead to stable and noise resistant q-bits using braiding of anyons in topological quantum computing.

In topological catalysis the semi-conducting and semi-metallic materials are characterized by robust metallic, conducting surface states, while being semiconducting in the bulk. Distinct from normal surfaces states, these states cannot be destroyed by chemical and physical measures. The electron mobility is huge and can be enhanced by an external magnetic field, thus ideal preconditions for an excellent catalyst.[42,47,48] Chiral Fermions have ultralong chiral surface arcs, so called Fermi arcs, and are excellent materials for hydrogen evolution reaction.[46]

## Author information

Corresponding Author:

†Nitesh Kumar (Nitesh.Kumar@cpfs.mpg.de)

*Claudia Felser (Claudia.Felser@cpfs.mpg.de)

## Notes
The authors declare no competing financial interest.

## Biographies

Nitesh Kumar is a group leader in the group of Prof. Claudia Felser at the Department of Solid State Chemistry, Max Planck Institute for Chemical Physics of Solids, Dresden, Germany. He obtained his M.S. (2010) and Ph.D. (2014) in Materials Science from Chemistry and Physics of Materials Unit (CPMU) at Jawaharlal Nehru Centre for Advanced Scientific Research (JNCASR), Bangalore under the combined supervision of Prof. C. N. R. Rao and Prof. A. Sundaresan. His research interests lie in single crystals growth and electronic properties of topological quantum materials.

Satya N. Guin received his M. Sc. (2011) in Chemistry from University of Kalyani and Ph.D. (2017) in Chemistry from the New Chemistry Unit (NCU), Jawaharlal Nehru Centre for Advanced Scientific Research (JNCASR), Bangalore under the guidance of Prof. Kanishka Biswas. He is currently a group leader with Prof. Claudia Felser at the Department of Solid State Chemistry, Max Planck Institute for Chemical Physics of Solids, Dresden, Germany. He is a recipient of Humboldt fellowship for postdoctoral researchers from Alexander von Humboldt foundation, Germany (2018). He is pursuing research in solid state chemistry, thermoelectrics, topological quantum materials and structure-property correlation.

Kaustuv Manna is a group leader in Max-Planck-Institute for Chemical Physics of Solids, Dresden, Germany in Prof. Claudia Felser's group. Before joining the institute, he pursued his first postdoctoral research in Leibniz Institute for Solid State and Materials Research Dresden IFW during May 2014 to June 2015 under the guidance of Dr. Sabine Wurmehl and Prof. Bernd Buchner. He obtained his Ph.D. degree from Department of Physics, Indian Institute of Science, India in 2014 under the guidance of Prof. P. S. Anil Kumar and Dr. Suja Elizabeth. He received his M. Sc. degree in Physics, from Department of Physics, Indian Institute of Technology (IIT), Guwahati, Assam, India in 2008. His research interests include extensive single crystal growth of various quantum materials including intermetallic alloys, oxides etc. and detailed structural, magnetic and transport characterizations.

Chandra Shekhar is a permanent senior group leader in Max Planck Institute for Chemical Physics of Solids, Dresden. Before joining this position, he was a postdoctoral scientist in Johannes Gutenberg University, Mainz and a guest scientist in IFW Dresden, Germany. He obtained his M.Sc. in Spectroscopy Physics, Ph.D. in Condensed Matter Physics from Banaras Hindu University and B. Sc. in both Physics and Chemistry from Purvanchal University, India. His multi specialized education played a crucial role to build up an early research career in the area of Physics and Chemistry of solids in Dresden.

Claudia Felser studied chemistry and physics at the University of Cologne, completing her doctorate in physical chemistry in 1994. She is currently Director at the Max Planck Institute for Chemical Physics of Solids in Dresden. She is a member of the Leopoldina, the German National Academy of Sciences, and an International Member of National Academy of Engineering, USA. In 2019 she received the APS James C. McGroddy Prize for New Materials together with Bernevig and Dai.

## Acknowledgements

The authors acknowledge financial support from the European Research Council (ERC) Advanced Grant No. 291472 "Idea Heusler" and 742068 "TOP-MAT"; European Union's Horizon 2020 research and innovation program (grant No. 824123 and 766566); Deutsche Forschungsgemeinschaft (DFG) through SFB 1143 and the Würzburg-Dresden Cluster of Excellence on Complexity and Topology in Quantum Matter– *ct.qmat* (EXC 2147, Project No. 390858490). S.N.G. thanks the Alexander von Humboldt Foundation for fellowship.

Table of contents

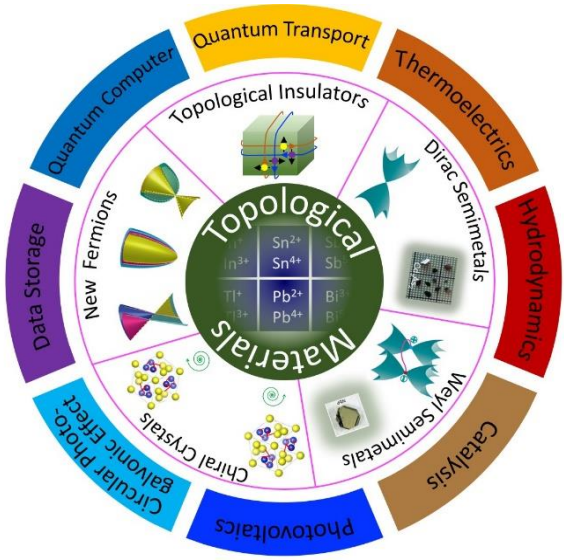